\documentclass[aps,prb,floatfix,showpacs,twocolumn]{revtex4}
\usepackage{amssymb,amsmath}
\usepackage{stmaryrd}
\usepackage{graphicx}
\usepackage{epsfig}
\usepackage{psfrag}
\newcommand{\bm}{\boldsymbol}

\begin{document}

\hsize\textwidth\columnwidth\hsize\csname@twocolumnfalse\endcsname
\title{Magneto-Optical Faraday and Kerr Effects 
in Topological Insulator Films \\
and in Other Layered Quantized Hall Systems} 

\author{Wang-Kong Tse}
\author{A.~H. MacDonald}
\affiliation{Department of Physics, University of Texas, Austin, Texas 78712, USA}

\begin{abstract}

We present a theory of the magneto-optical Faraday and Kerr effects
of topological insulator (TI) films.   For film thicknesses
short compared to wavelength, we find that the low-frequency Faraday
effect in ideal systems is quantized at integer multiples of the fine structure 
constant, and that the Kerr effect exhibits a giant $\pi/2$ rotation for either  
normal or oblique incidence.  For
thick films that contain an integer number of half wavelengths, we find
that the Faraday and Kerr effects are both quantized at 
integer multiples of the fine structure constant.  For TI films with
bulk parallel conduction, we obtain a criterion  
for the observability of surface-dominated magneto-optical effects. 
For thin samples supported by a substrate, 
we find that the 
universal Faraday and Kerr effects are present when 
the substrate is thin compared to the optical wavelength or  
when the frequency matches a thick-substrate cavity resonance.
Our theory applies equally well to any system with two
conducting layers that exhibit quantum Hall effects. 
\end{abstract}
\maketitle

\newpage



\section{Introduction}

Because topological insulators (TIs) have gapless helical surface states
\cite{FuKaneMele,MooreBalents,Roy,KaneHasan} that respond strongly to 
time-reversal symmetry breaking perturbations, magneto-optical studies have emerged
as an important tool for their characterization\cite{Drew,Armitage,Molenkamp,SCZhangRev,Vanderbilt,Tse_1,Tse_2,HasanPhysics,QiBulkSubstrate,Tkachov}. 
This paper expands on previous work\cite{Tse_1,Tse_2} in which 
we demonstrated that ideal TIs exhibit striking unversal features in their long-wavelength response --- 
a universal Faraday angle equal to the fine structure
constant and a giant $90$ degrees Kerr rotation. 
The present paper details the formalism used to obtain these results
and generalizes the theory to new circumstances motivated by current experimental activity.
In particular, we include the influence of bulk conduction on the magneto-optical properties,
analyze the role of the substrate material, and examine the case of oblique incidence of the light source.
We also extend our theory to include
thick TI films in which the electromagnetic wave can excite one of the
cavity resonance modes of the film.  Although we focus on the case of TI thin films,
our results apply generically to systems containing two conducting layers that exhibit quantum
Hall effects.

When bulk conduction and surface longitudinal 
conduction are both negligible, the magneto-optical properties of a TI thin film
can be elegantly characterized by adding a magneto-electric 
coupling term to the electromagnetic Lagrangian to obtain {\em
  topological field theory} \cite{SCZhangRev}.  
This description of magneto-electric properties shows that TIs  provide a solid-state realization of axion electrodynamics, similar to those anticipated by Wilzcek in Ref.~[\onlinecite{Wilczek}].
The topological field theory formulation of magneto-electric properties can be derived by integrating out the electronic 
degrees of freedom to obtain the magneto-electric polarizability of the bulk insulator, 
which is expressible as a Chern-Simons $3$ form \cite{SCZhangRev,Vanderbilt}. 
By appealing to bulk time-reversal invariance considerations, it is possible to conclude that the 
coupling constant $\theta$ of the topological field theory\cite{SCZhangRev,Vanderbilt} is 
$0\,\mathrm{mod}(2\pi)$ for ordinary insulators and $\pi\,\mathrm{mod}(2\pi)$ for TIs. 
These possibilities correspond respectively to integer quantized surface Hall conductances 
in the case of ordinary insulators and to half-integer quantized
surface Hall conductances \cite{Tse_2} 
in the case 
of topological insulators, providing a demonstration of this important TI property.
The topological field theory approach has a number of limitations, however, in describing real experiments
because i) it does not account for the surface longitudinal conductance which is never precisely zero at finite temperatures 
even when the quantum Hall effect is well established, ii) the 
surface Hall conductivity in topological field theory is ambiguous up
to an integer multiple of 
$e^2/h$, and iii) real thin film samples often have a finite bulk conductivity that is not 
readily incorporated.  In addition, the thin film geometry normally used for magneto-optical 
studies requires seemingly artificial spatial profiles of the $\theta$ coupling constant,
as discussed in the following paragraph. 
For these reasons we prefer to model the surface Hall conductivities using a 
microscopic two-dimensional massless Dirac model for the TI surface states 
that is fully detailed below. 
The main disadvantage of our 
approach is that it captures the precise quantization of the {\em DC} surface Hall conductivity 
only when the massless Dirac model's ultraviolet cutoff is set to
infinity. An important lesson from our approach is that topological field theory applies only 
when time-reversal symmetry breaking is strong enough to overcome disorder and establish 
a surface quantum Hall effect, and then only in the limit of temperatures and 
frequencies small compared to the surface gap induced by time-reversal symmetry breaking. 

\begin{figure*}
  \includegraphics[width=18.5cm,angle=0]{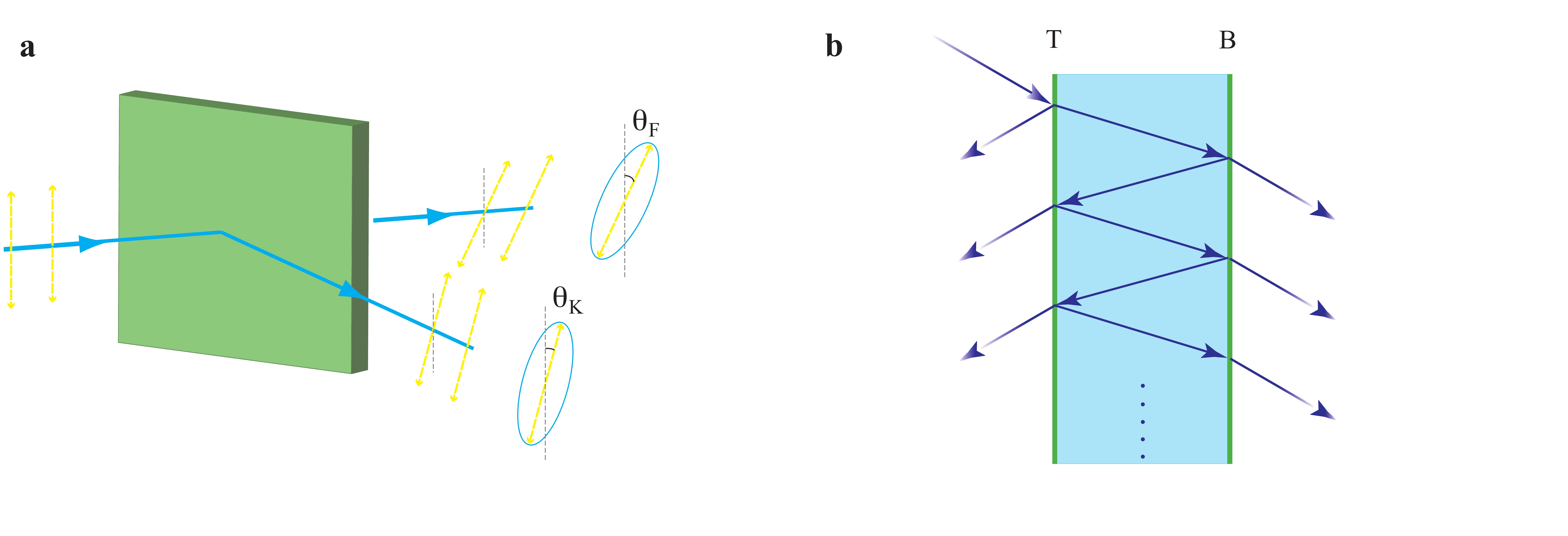}
\caption{(Color online) (a). Schematic illustration of the Faraday and Kerr effects. Incident linearly polarized light becomes elliptically polarized after transmission (Faraday effect) and reflection (Kerr effect), with polarization plane angle rotations $\theta_{\mathrm{F}}$ and $\theta_{\mathrm{K}}$ respectively. (b) Fabry-Perot-like reflection and transmission in the TI film geometry. `T' stands for the top surface and `B' the bottom surface.}
\label{TIslab}
\end{figure*}

The original discussion of Wilzcek \cite{Wilczek} imagined axion electrodynamics 
induced by a bulk spherical medium with a non-zero $\theta$ parameter
separated from vacuum with $\theta=0$ by a single simply-connected 
surface.  In the case of a spherical TI sample, this formulation can
correctly capture the material's surface Hall conductivity.  
Similarly, the case of an ideal semi-infinite TI slab \cite{SCZhangRev}
can be described by a topological field theory model with $\theta \ne 0$ in the 
topological insulator and $\theta=0$ in vacuum.  
In a magneto-optics setting, however, a propagating electromagnetic
wave necessarily interacts with two nearby surfaces,
since a real TI thin film sample has both top and bottom surfaces.
Assuming that the mechanism that breaks time-reversal invariance at the TI 
surface does so in the same sense on both top and bottom, both surfaces will
have the same Hall conductivity.  In topological field theory, this would be 
captured by a model in which the discontinuity in $\theta$ in the direction
of light propagation has the same sign and magnitude at both surfaces.
To describe a thin film with identical half-quantized Hall conductivities on opposite surfaces 
it is then necessary to take $\theta = \pm 2\pi$ in one of the vacuum regions.
A TI thin film model in which the surrounding material has $\theta =0$
everywhere would instead describe a system with opposite Hall conductivities on opposite 
surfaces.  Because of the opposite contributions, the magneto-optical effects would then vanish for films thinner than a 
wavelength. 

Given the surface massless Dirac models, our approach to magneto-electric properties 
is more conventional.  Magneto-electric properties are completely determined by 
the conductivity of the material, including bulk and surface, longitudinal and Hall 
contributions.  The role of the theta-term in axion electrodynamics equations is completely
replaced by the appearance of explicit surface Hall conductivities that 
influence electromagnetic wave boundary conditions.
These calculations make it clear that magneto-electric properties depend essentially on the numerical values 
of the surface conductivities, not just on whether they are quantized or half-quantized.
This is especially important when surface time-reversal invariance is broken 
by an external magnetic field, since both the sign and 
the magnitude of the surface currents will be sensitive to the position of the 
Fermi level within the bulk gap.  In addition to this advantage, the optical
response of the surface Dirac fermions can be evaluated 
microscopically, enabling a natural incorporation of dynamical and
many-body effects.   




Interesting magneto-optical effects occur in both low-frequency and
higher frequency regimes. In the latter case, we have found
interband absorption \cite{Tse_1} and cyclotron resonance \cite{Tse_2}
features in the Faraday and Kerr rotations that are dramatically
enhanced by the cavity confinement effect of the TI thin film.
We focus mainly on the large topological magneto-optical effects at low frequencies 
which can be observed only if the following conditions are satisfied:
(1) The TI surface has a quantized Hall effect allowed by 
time-reversal breaking due to either external magnetic field 
or exchange coupling to external electronic degrees of freedom.
We know from vast experience with the quantum Hall effect in an
external magnetic field in graphene, which is also described by
a massless Dirac equation, that the quantum Hall effect can
occur at quite weak magnetic fields when the Fermi level is 
close to the Dirac point and disorder is very weak.  Exchange-coupling 
to spin yields a half-quantized anomalous Hall effect in the absence of 
disorder \cite{Haldane,QSH_Kane}. Although there are no experimental examples of quantized 
anomalous Hall effect as yet, the requirements on time-reversal breaking 
perturbation strength in the presence of disorder are likely to be similar.
The magneto-electric anomalies require large Hall effects at 
finite electromagnetic wave frequency. This condition requires that 
the frequency be much smaller than the surface gap. 
(2) The dimension of the TI along the 
direction of light travel should be shorter than the electromagnetic wavelength \textit{or} 
a nonzero integer multiple of the half wavelength. In either case, the
dielectric properties of the TI bulk medium that separates the two quantized
Hall layers do not influence the transmitted and reflected light.  
When these conditions are satisfied, the magneto-optical effects are
universal and topologically protected against weak surface disorder.

The outline of our paper is as follows. In the Section II, we review linear
response theory for the optical conductivity tensor of TI surface
helical quasiparticles. In Section III, we summarize the
electromagnetic scattering formalism appropriate for a system
with two metallic surfaces surrounded by dielectrics, 
applicable to TI films and to other similar layered systems. 
We then present and discuss our
results for the magneto-optical Faraday 
and Kerr effects, first for TI films that are thinner than a 
wavelength and then for the general case. 
In Section V we address some issues pertinent to experiments
including the influence of bulk conduction,
the role of oblique incidence, and the role of substrates.
Finally in Section VI we
discuss the closely related magneto-optical properties
of graphene-based layered massless Dirac systems.





\section{Dynamical Response of a Spin-Helical Dirac Fermions}  

We first derive explicit analytic expressions for the
longitudinal and Hall optical conductivities of the spin-helical quasiparticles of topological insulator
surfaces with time-reversal symmetry broken in two different ways: (1) an exchange field that couples to 
spin, and (2) an external magnetic field that couples to both spin and
orbital degrees of freedom.
The former case can be realized 
by exchange coupling between the TI surface
and an adjacent ferromagnetic insulator \cite{Tse_1,SCZhangRev}. 
The response of TI surface carriers to exchange coupling 
is unique, and ongoing progress along this direction has been reported
by several experimental groups
\cite{Hasan_doping1,Hasan_doping2,Shen_doping,Cui_doping}. 
%

In the presence of time-reversal symmetry breaking, the massless Dirac 
Hamiltonians for the top (T) and bottom (B) surfaces are 
\begin{equation}
H = (-1)^L \; [ v\bm{\tau}\cdot(-i\nabla+e\bm{A}/c) + V/2] +
\Delta\tau_z,
\label{Ham}
\end{equation}
%
where $\bm{\tau}$ is the spin Pauli matrix vector (expressed in $90^{\circ}$ rotated basis from the real spins \cite{spindirection}), $\bm{A}$ 
is the magnetic vector potential, $\Delta$ is the Zeeman coupling strength, $V$ accounts for a possible 
potential difference between top and bottom surfaces due to doping or external gates, and $L = 0,1$ 
for the top ($0$) and bottom ($1$) surfaces. Note that in spite of the
sign difference in the kinetic energy terms for the top and bottom
surfaces, the conductivities are identical on the two
surfaces. The massless Dirac surface
states description is valid for energies below the energy cut-off of the Dirac
Hamiltonian $\varepsilon_{\mathrm{c}}$, which we associate with the separation between the Dirac
point and the closest bulk band. 


\subsection{Exchange Field}

Time-reversal symmetry breaking by an exchange field can be realized by
interfacing the TI surface with an insulating ferromagnet with magnetization oriented 
perpendicularly.  Magnetic 
proximity coupling with strength $\Delta$ will favor alignment of the TI surface spins.
The numerical value of $\Delta$ would be determined
by the orbital hybridization between the ferromagnetic material and the TI
surface. $A = 0$ in Eq.~(\ref{Ham}) in the absence of an applied
magnetic field. In this limit, the Dirac cone is gapped with conduction ($\mu = 1$) and valance ($\mu
= -1$) band dispersions
and eigenstates 
%
\begin{equation}
\varepsilon_{k\mu} = \mu\varepsilon_k,\,\,\,\,\,
\vert k\mu\rangle = \begin{bmatrix}
C_{\uparrow k\mu}\\ C_{\downarrow k\mu} e^{i\phi_k}\end{bmatrix}. 
\label{EnergyEx}
\end{equation} 
where $\varepsilon_k = \sqrt{(vk)^2+\Delta^2}$, $\phi_k$ is the azimuthal angle of the crystal momentum, and
\begin{eqnarray}
C_{\uparrow k\mu} &=&
\mathrm{sgn}(\mu)\sqrt{\varepsilon_k+\mathrm{sgn}(\mu)\Delta}/\sqrt{2\varepsilon_k},
\nonumber \\
C_{\downarrow k\mu} &=& \sqrt{\varepsilon_k-\mathrm{sgn}(\mu)\Delta}/\sqrt{2\varepsilon_k},
\label{CEx}
\end{eqnarray}
%

The optical conductivity tensor of the helical quasiparticles on topological
insulator surface can be obtained from the Kubo formula:
\begin{equation}
\sigma_{\alpha\beta}\left(\omega\right) = ig\sum_{k}\sum_{\mu\mu'}
\frac{f_{k\mu}-f_{k\mu'}}{\varepsilon_{k\mu}-\varepsilon_{k\mu'}}\frac{\langle k\mu \vert
  j_{\alpha} \vert k\mu' \rangle \langle k\mu' \vert
  j_{\beta} \vert k\mu
  \rangle}{\omega+\varepsilon_{k\mu}-\varepsilon_{k\mu'}+i/2\tau_{\mathrm{s}}},
\label{KuboExchange}
\end{equation}
where $\alpha,\beta = \{x,y\}$, $\mu,\mu' = \pm 1$ denote the band index, $f_{k\mu}$ is the
Fermi factor for band $\mu$, $1/\tau_{\mathrm{s}}$ is the 
quasiparticle lifetime broadening of the surface states, and $g$ is the (odd) number of
Dirac cones on the TI surface, which we take for simplicity and
concreteness to be $1$.  This expression is not quantitatively correct when disorder 
is present ({\em i.e.} when $\tau_{\mathrm{s}}$ is finite), because it does not capture the localization physics that is important in the 
quantum Hall regime, but it is adequate for our present interest.  We evaluate 
the longitudinal conductivity $\sigma_{xx}(\omega) = \sigma_{xx}^{\mathcal{R}}+i\sigma_{xx}^{\mathcal{I}}$ and the 
Hall conductivity $\sigma_{xy}(\omega) =
\sigma_{xy}^{\mathcal{R}}+i\sigma_{xy}^{\mathcal{I}}$ in the
topological transport regime when the Fermi level lies within the TRS breaking gap. 
Since all surface optical conductivities eventually appear in the outgoing 
electromagnetic fields in the combination $\sigma_{\alpha\beta}/c$, we
shall adopt the `natural units' for the optical conductivities and express ${\sigma}_{\alpha\beta}$ in $e^2/\hbar = \alpha \,
c$ units ($\alpha = 1/137$ is the vacuum fine structure constant) and set $c = 1$ except where specified. 
For the dissipative components of the optical conductivity we find that:  
\begin{eqnarray}
\sigma_{xx}^{\mathcal{R}} &=& \frac{1}{16\pi
  x(\omega^2+\Gamma^2)^2}\nonumber \\
&&\times\left\{x\left[(\omega^2+\Gamma^2)^2+4(\omega^2-\Gamma^2)\Delta^2\right]g(x)\right.
  \nonumber \\
&&\left.+4\Gamma\Delta^2\left[\omega^2+\Gamma^2+\omega
    xf(x)\right]\right\}\big\vert_{x
    = \Delta}^{x = \varepsilon_{\mathrm{c}}} \nonumber \\
\sigma_{xy}^{\mathcal{I}} &=&
\frac{\Delta}{8\pi(\omega^2+\Gamma^2)} \left[-2\omega g(x)-\Gamma f(x)\right]\big\vert_{x
    = \Delta}^{x = \varepsilon_{\mathrm{c}}} 
\label{Rsxx}
\end{eqnarray}
For the reactive, non-dissipative components of the optical conductivity $\sigma_{xx}^{\mathcal{I}}$ and $\sigma_{xy}^{\mathcal{R}}$, which are due to off-shell virtual transitions, we find that: 
\begin{eqnarray}
\sigma_{xx}^{\mathcal{I}} &=&
\frac{1}{32\pi x(\omega^2+\Gamma^2)^2}\left\{8\omega\Delta^2(\omega^2+\Gamma^2)
-16x\omega\Gamma\Delta^2 g(x) \right.
\nonumber \\
&&\left.+x\left[(\omega^2+\Gamma^2)^2+4\Delta^2
    (\omega^2-\Gamma^2)\right] f(x)\right\}\big\vert_{x = \Delta}^{x = \varepsilon_{\mathrm{c}}} \nonumber \\
\sigma_{xy}^{\mathcal{R}} &=& \frac{\Delta}{8\pi(\omega^2+\Gamma^2)}
\left[-2\Gamma g(x)+\omega f(x)\right]\big\vert_{x
    = \Delta}^{x = \varepsilon_{\mathrm{c}}} \nonumber \\
\label{Isxx}
\end{eqnarray}
where $R(x)\vert_{x=x_1}^{x = x_2}$ means $R(x_2)-R(x_1)$, $\Gamma = 1/2\tau_{\mathrm{s}}$, and
\begin{eqnarray}
f(x) &=&
\mathrm{ln}\left\vert\frac{(\omega-2x)^2+\Gamma^2}{(\omega+2x)^2+\Gamma^2}\right\vert,
\nonumber \\
g(x) &=&
\tan^{-1}\left(\frac{\omega-2x}{\Gamma}\right)-\tan^{-1}\left(\frac{\omega+2x}{\Gamma}\right).
\label{fg}
\end{eqnarray}
When the disorder broadening is small such that $\Gamma \ll \Delta$,
it is useful to obtain analytic results in the disorder-free limit
$\Gamma \to 0$, in which case 
\begin{equation}
g(x)\big\vert_{x = \Delta}^{x = \varepsilon_{\mathrm{c}}} =
-\pi\theta(\vert\omega\vert-2\vert\Delta\vert), \label{gG0}
\end{equation}
and Eqs.~(\ref{Rsxx})-(\ref{fg}) reduce to our previous 
results [Eqs.~(2)-(3) in Ref.~\onlinecite{Tse_1}] obtained from the quantum kinetic equation approach.
%

\subsection{External Quantizing Magnetic Field}


Landau level (LL) quantization of the TI's surface Dirac cones has recently been
observed by STM experiments \cite{STM1,STM2}. In the presence of a quantizing field, 
the vector potential in Eq.~(\ref{Ham}) is given in the Landau  
gauge by $\bm{A} = (0,Bx)$ and the Zeeman coupling by $\Delta =
g_{\mathrm{J}}\mu_{\mathrm{B}}B/2$, where $g_{\mathrm{J}}$ is the
electron g factor.
Define raising and lowering operators $a = 
(\ell_B/\sqrt{2})[\partial_x+(x+x_0)/\ell_B^2]$ and $a^{\dagger} = 
(\ell_B/\sqrt{2})[-\partial_x+(x+x_0)/\ell_B^2]$ with $\ell_B = 
1/\sqrt{e\vert B\vert}$ the magnetic length and $x_0 = k_y \ell_B^2$ 
the guiding center coordinate, 
Eq.~(\ref{Ham}) can be written as 
%
\begin{equation}
\begin{bmatrix}
{ \Delta} & -i\left({\sqrt{2}v}/{\ell_B}\right) a \\
i\left({\sqrt{2}v}/{\ell_B}\right) a^{\dagger} & -{\Delta} \end{bmatrix}
\vert \overline{n} \rangle = \varepsilon \vert \overline{n} \rangle,
\label{eigenenergy}
\end{equation}
where $\vert \overline{\cdots}\rangle$ denotes an eigenspinor. 
The LLs are labeled by integers $n$ and for $n \ne 0$ and have eigenenergies [relative to the
respective Dirac point energies $(-1)^LV/2$] 
and eigenspinors
%
\begin{eqnarray}
\varepsilon_n = \mathrm{sgn}(n)\sqrt{\frac{2v^2}{\ell_B^2}\vert n
  \vert+\Delta^2},\,\,\,\,\,\vert \overline{n} \rangle = 
\begin{bmatrix}
-i\mathcal{C}_{\uparrow n}\vert \vert n \vert -1\rangle \\
\mathcal{C}_{\downarrow n}\vert \vert n \vert\rangle
\end{bmatrix}. 
\label{LL}
\end{eqnarray}
%
where $\vert\vert n \vert \rangle$ without an overbar denotes a Fock state ($\vert n \vert$ is the
absolute value of $n$), 
and 
\begin{eqnarray}
\mathcal{C}_{\uparrow n} &=& \mathrm{sgn}(n)\sqrt{\varepsilon_{\vert n
    \vert}+\mathrm{sgn}(n)\Delta}/\sqrt{2\varepsilon_{\vert n    \vert}},
\nonumber \\
\mathcal{C}_{\downarrow n} &=& \sqrt{\varepsilon_{\vert n
    \vert}-\mathrm{sgn}(n)\Delta}/\sqrt{2\varepsilon_{\vert n
    \vert}}. \label{CupCdown}
\end{eqnarray}
%
In the $n=0$ LL spins are aligned with the perpendicular field so that 
%
\begin{eqnarray}
\varepsilon_0 = -\Delta,\,\,\,\,\,\vert \overline{0} \rangle = 
\begin{bmatrix}
0 \\
\vert 0 \rangle
\end{bmatrix}.
\label{LL0}
\end{eqnarray}
%

In the quantum Hall regime ($\Omega_B\tau_{\mathrm{s}}
\gg 1$ where 
$\Omega_B = v/\ell_B$ is a characteristic frequency typical of the LLs
spacing), the conductivity tensor can be expressed in LL basis as 
\begin{equation}
\sigma_{\alpha\beta}\left(\omega\right) = \frac{ig}{2\pi
  \ell_B^2}\sum_{nn'}
\frac{f_n-f_{n'}}{\varepsilon_n-\varepsilon_{n'}}\frac{\langle \overline{n} \vert
  j_{\alpha} \vert \overline{n'} \rangle \langle \overline{n'} \vert
  j_{\beta} \vert \overline{n}
  \rangle}{\omega+\varepsilon_n-\varepsilon_{n'}+i/2\tau_{\mathrm{s}}},
\label{KuboLL}
\end{equation}
where the current operator is $\bm{j} = ie\left[H,\bm{x}\right] =
ev\bm{\tau}$. For convenience we rewrite the LL index as $n = sm$, where $m = 0, 1,
2, \cdots N_{\mathrm{c}}$ and $s = \pm 1$ for electron-like and hole-like LLs.
The current matrix element $\langle \overline{n} \vert
  j_{\alpha} \vert \overline{n'} \rangle$ captures the selection
  rule $\vert n'\vert-\vert n\vert =\pm 1$ for LL transitions.  
  After some algebra we find that the
  conductivity tensor Eq.~(\ref{KuboLL}) in 
$e^2/\hbar = \alpha c $ units is given by
\begin{eqnarray}
\sigma_{\alpha\beta}(\omega) = \frac{v^2}{2\pi \ell_B^2}\sum_{m = 0}^{N_{\mathrm{c}}-1}\sum_{s,s' = \pm 1}
\frac{f_{sm}-f_{s'(m+1)}}{\varepsilon_{sm}-\varepsilon_{s'(m+1)}} \;
\Gamma_{\alpha\beta}^{s,s'}(m,\omega), \nonumber \\
\label{Cond} 
\end{eqnarray}
%
where 
\begin{eqnarray}
&&\Gamma_{\left\{\begin{subarray}{l} xx \\ xy \end{subarray}\right\}}^{s,s'}(m,\omega) = 
-\left\{\begin{array}{c} i \\ 1 \end{array}
\right\}\mathcal{C}_{\uparrow s'(m+1)}^2\mathcal{C}_{\downarrow
  sm}^2 
\label{function_f} \\
&&\left(\frac{1}{\omega-\varepsilon_{sm}+\varepsilon_{s'(m+1)}+i/2\tau_{\mathrm{s}}}\pm\frac{1}{\omega+\varepsilon_{sm}-\varepsilon_{s'(m+1)}+i/2\tau_{\mathrm{s}}}\right). \nonumber
\end{eqnarray}
In Eq.(~\ref{function_f}), $N_{\mathrm{c}} \simeq \ell_B^2(\varepsilon_{\mathrm{c}}^2-\Delta^2)/2v^2$ 
is the largest LL index with an energy smaller than the ultraviolet
cut-off $\varepsilon_{\mathrm{c}}$, prefactor $i$ and sign `$+$'
inside the parenthesis apply to the 
$\Gamma_{xx}$ expression, $1$ and `$-$' to the $\Gamma_{xy}$ expression. 
Eqs.~(\ref{Cond})-(\ref{function_f}) express 
$\sigma_{\alpha\beta}$ as a sum over interband and intraband dipole-allowed 
transitions which satisfy $\vert n'\vert-\vert n\vert =\pm 1$.
In the $\omega=T=\tau_{\mathrm{s}}^{-1}=0$ limit Eq.~(\ref{Cond}) yields 
correct half-quantized plateau values for the Hall conductivity.

\section{Light Propagation through a TI Slab}

In this section, we formulate the problem of electromagnetic wave
scattering through a topological insulator slab illustrated schematically in
Fig.~\ref{TIslab}(b).  Here we discuss only the normal incidence case.  More 
general results for the oblique incidence case are presented in an Appendix. 

Consider an electromagnetic wave propagating along the $z$ direction through two materials, labeled by $i$ and $j$, with dielectric constant and magnetic permeability $\epsilon_i,\mu_i$ and $\epsilon_j,\mu_j$ respectively. The interface between them is at $z = a_i$.  We write the electric field component of the electromagnetic field in the form 
\begin{eqnarray} 
\tilde{E}^i = 
e^{ik_iz}\left[\begin{array}{c} E_x^{ti} \\ E_y^{ti}\end{array}\right]
+e^{-ik_iz}\left[\begin{array}{c} E_x^{ri} \\ E_y^{ri}\end{array}\right], \label{S17}
\end{eqnarray}
where the tilde accents denote vectors $\tilde{E} = [E_x\,\,\,E_y]^{\mathrm{T}}$, the superscripts `$\mathrm{r}$' and `$\mathrm{t}$' on
$\tilde{E}$ denote the reflected and transmitted components of the
electric field, and $k_{i} = (\omega/c)\sqrt{\epsilon_{i}\mu_{i}}$ is the
wavevector in medium $i$.
The 
corresponding magnetic field is given by Faraday's law as 
%
\begin{eqnarray}
\tilde{H}^i = \sqrt{\frac{\epsilon_i}{\mu_i}}\left\{
e^{ik_iz}\left[\begin{array}{c} -E_y^{ti} \\
    E_x^{ti} \end{array}\right] +e^{-ik_iz}\left[\begin{array}{c} E_y^{ri} \\
-E_x^{ri}\end{array}\right]\right\}, \label{S18}
\end{eqnarray}
The electric and magnetic fields at the interface $z = a_i$ satisfy
the electrodynamic boundary conditions $\tilde{E}^i = \tilde{E}^{j}$
and $-i\tau_y(\tilde{H}^j-\tilde{H}^i) = (4\pi/c)\tilde{J}^{i}$, 
where $\tau_y$ is the Pauli matrix and $\tilde{J}^i = \bar{\sigma}^i\tilde{E}^i$ is the surface current density at $z = a_i$. 
Note that this surface current can have longitudinal and Hall response components in this microscopic 
theory and not only Hall components as assumed in topological field theory.   

The scattering matrix that relates incoming $[\tilde{E}^{\mathrm{t}i}\,\,\,\tilde{E}^{\mathrm{r}j}]^{\mathrm{T}}$
and outgoing $[\tilde{E}^{\mathrm{r}i}\,\,\,\tilde{E}^{\mathrm{t}j}]^{\mathrm{T}}$ fields at a 
conducting interface can be written in the form
\begin{equation}
S = \left[\begin{array}{cc}
\bar{r} & \bar{t'} \\
\bar{t} & \bar{r'}\end{array}\right],
\label{Scat}
\end{equation}
where the superscripts `$\mathrm{r}$' and `$\mathrm{t}$' on $\tilde{E}$ 
denote reflected and transmitted components of the electric fields, and $\bar{r},\bar{r^{'}}$ and $\bar{t},\bar{t^{'}}$ are 
$2\times 2$ reflection and transmission tensors, which are of the form:
\begin{equation}
\bar{r} = \begin{bmatrix}
r_{xx} & r_{xy} \\
-r_{xy} & r_{yy} \end{bmatrix},\,\,\,\,\,
\bar{t} = \begin{bmatrix}
t_{xx} & t_{xy} \\
-t_{xy} & t_{yy} \end{bmatrix},
\label{RT}
\end{equation}
and similarly for $\bar{r^{'}}, \bar{t^{'}}$. Matching boundary conditions, we obtain
the following expressions for $\bar{r}, \bar{t}$ $\bar{r}^{'}$ and $\bar{t}^{'}$:
\begin{eqnarray}
&&\left\{\begin{array}{c}
r_{xx} \\
r_{xy}\end{array}\right\}
 = \frac{e^{i2k_ia_i}}{(\sqrt{\epsilon_i/\mu_i}+\sqrt{\epsilon_j/\mu_j}+4\pi{\sigma}_{xx})^2+(4\pi{\sigma}_{xy})^2} \nonumber \\
&&\times\left\{\begin{array}{c}
\epsilon_i/\mu_i-(\sqrt{\epsilon_j/\mu_j}+4\pi{\sigma}_{xx})^2-(4\pi{\sigma}_{xy})^2 \\
-8\pi \sqrt{\epsilon_i/\mu_i}{\sigma}_{xy}\end{array}\right\}, \label{normalr} 
\end{eqnarray}
\begin{eqnarray}
&&\left\{\begin{array}{c}
t_{xx} \\
t_{xy}\end{array}\right\}
= \frac{e^{i(k_i-k_j)a_i}}{(\sqrt{\epsilon_i/\mu_i}+\sqrt{\epsilon_j/\mu_j}+4\pi{\sigma}_{xx})^2+(4\pi{\sigma}_{xy})^2} \nonumber \\
&&\times\left\{\begin{array}{c}
2\sqrt{\epsilon_i/\mu_i}(\sqrt{\epsilon_i/\mu_i}+\sqrt{\epsilon_j/\mu_j}+4\pi{\sigma}_{xx}) \\
-8\pi\sqrt{\epsilon_i/\mu_i}{\sigma}_{xy}\end{array}\right\}. \label{normalt} 
\end{eqnarray}
%
For normal incidence, the two diagonal elements of the reflection and transmission
matrices are identical: $r_{yy}^{(')} = r_{xx}^{(')}$ and $t_{yy}^{(')}
= t_{xx}^{(')}$. $\bar{r'}$ can be obtained from $\bar{r}$ by 
making the replacement $k_i \to -k_j$ and
interchanging $\epsilon_i/\mu_i$ and $\epsilon_j/\mu_j$, and $\bar{t'}$
from $\bar{t}$ by interchanging $\epsilon_i/\mu_i$ and
$\epsilon_j/\mu_j$. In addition, $\bar{t}$ and $\bar{t}'$ are related by 
$\bar{t}/\sqrt{\epsilon_i/\mu_i} = \bar{t}'/\sqrt{\epsilon_j/\mu_j}$.

Scattering from a TI film presents an electromagnetic problem in which scattering occurs from two interfaces 
at which currents flow and dielectric constants are discontinuous. 
The reflection and transmission tensor
can be composed from the single-interface scattering matrices
$\bar{r^{(')}}, \bar{t^{(')}}$ for the top and bottom surfaces to obtain
\begin{eqnarray}
\bar{r} &=&
\bar{r_{\mathrm{T}}}+\bar{t_{\mathrm{T}}'}\bar{r_{\mathrm{B}}}\left(\bm{1}-\bar{r_{\mathrm{T}}'}\bar{r_{\mathrm{B}}}\right)^{-1}\bar{t_{\mathrm{T}}}, \label{r20} \\
\bar{t} &=&
\bar{t_{\mathrm{B}}}\left(\bm{1}-\bar{r_{\mathrm{T}}'}\bar{r_{\mathrm{B}}}\right)^{-1}\bar{t_{\mathrm{T}}}. \label{t20}
\end{eqnarray}
The presence of a dielectric substrate underneath the TI film is easily accounted for by 
propagating the reflection and transmission tensors
Eqs.~(\ref{r20})-(\ref{t20}) through an additional layer of
dielectric. Detailed expressions for the reflection and
transmission tensors that allow for oblique incidence and account for a 
dielectric substrate are given in an Appendix.  

\section{Magneto-optical Faraday and Kerr Effects}

For linearly polarized incoming light, the Faraday and Kerr angles can be defined in terms of the relative rotations 
of left-handed and right-handed circularly polarized light to obtain: 
\begin{eqnarray}
\theta_{\mathrm{F}} &=&
\left(\mathrm{arg}\{E_{+}^{\mathrm{t}}\}-\mathrm{arg}\{E_{-}^{\mathrm{t}}\}\right)/2, \label{FaraDef} \\
\theta_{\mathrm{K}} &=&
\left(\mathrm{arg}\{E_{+}^{\mathrm{r}}\}-\mathrm{arg}\{E_{-}^{\mathrm{r}}\}\right)/2, \label{KerrDef}
\end{eqnarray}
where ${E}_{\pm}^{\mathrm{r,t}} = {E}_x ^{\mathrm{r,t}}\pm i{E}_y
^{\mathrm{r,t}}$ are the left-handed (+) and right-handed (-)
circularly polarized 
components of the outgoing electric fields.

In this section, we present our results for an ideal topological insulator under 
normal light incidence. We then discuss non-ideal effects that may be
relevant in experimental situations in Section V. It is important to 
emphasize that the magneto-optical effects are essentially the same 
in this limit for time-reversal symmetry broken by exchange coupling 
or by a quantizing
magnetic field.  In the case of a quantizing field, there are many gaps 
in the surface spectrum because of Landau quantization.  
The quantized Hall conductivity in $e^2/h$ units is equal to the 
filling factor $\nu_{\mathrm{T,B}}$.  
The largest gap in the magnetic field case occurs occurs at $\nu_{\mathrm{T,B}}=1/2$ 
and has the same Hall conductivity as for the Zeeman gap case.
In the magnetic field case, it 
is possible to shift the Hall conductivities of either surface by 
integer multiples of $e^2/h$ away from $e^2/2h$
simply by shifting the position of the Dirac point relative to the 
chemical potential, and this shift would influence the magneto-optical effects.
When the chemical potential is placed in the largest gap in the magnetic field case, 
the only differences between the two scenarios are in the details of the higher
frequency response. 

\subsection{Thin Film $d \ll \lambda$}


First we consider the case of a TI film that is thinner than the light wavelength. 
In this limit it follows from Faraday's law that the electric field is spatially constant across the film
so that the two interfaces can be considered as one.
Amp\`ere's Law implies that the magnetic field changes 
by a value proportional to the current 
integrated across the TI film,
\begin{equation}
-i\tau_y(\tilde{H}^{\mathrm{T}}-\tilde{H}^{\mathrm{B}}) = (4\pi/c)\left(\bar{\sigma}_{\mathrm{T}}+\bar{\sigma}_{\mathrm{B}}\right)\tilde{E},
\label{LongWave}
\end{equation}
where $\tilde{H}^{\mathrm{T,B}}$ denotes the magnetic fields in the
top and bottom vacuum regions outside of the film.  
Eq.~(\ref{LongWave}) says that, from the viewpoint of the long
electromagnetic wave, the TI film behaves effectively as a
single two-dimensional surface with a conductivity equal to the
conductivities integrated across the film. 
We therefore obtain the transmitted and reflected fields
\begin{eqnarray}
\tilde{E}^{\mathrm{t}} &=& \frac{1}{\left(2+4\pi\sigma_{xx}\right)^2+\left(4\pi\sigma_{xy}\right)^2}\begin{bmatrix} 4\left(1+2\pi\sigma_{xx}\right) \\
8\pi\sigma_{xy}\end{bmatrix}, \label{T_Efield} \\
%
\tilde{E}^{\mathrm{r}} &=& 
\frac{1}{\left(2+4\pi\sigma_{xx}\right)^2+\left(4\pi\sigma_{xy}\right)^2} \nonumber \\
&&\begin{bmatrix} 1-\left(1+4\pi\sigma_{xx}\right)^2-\left(4\pi\sigma_{xy}\right)^2 \\
8\pi\sigma_{xy}\end{bmatrix}, \label{R_Efield}
\end{eqnarray}
for simplicity here we use $\sigma_{xx}, \sigma_{xy}$ to denote the
total longitudinal and Hall conductivities from both surfaces, respectively. An important observation is
that in this limit the transmitted and reflected fields are independent of the bulk dielectric properties of
the TI film.  For weak disorder and frequencies much smaller than characteristic transition frequencies 
($\omega \ll \Delta$ for the exchange field case and $\omega \ll \Omega_B $ for the magnetic field case), 
the optical conductivity has only a dissipationless \textit{DC} Hall conductivity contribution:
\begin{equation}
\sigma_{xy}^{\mathcal{R}} = \frac{\nu_{\mathrm{T,B}}}{2\pi}, 
%
\label{Hall}
\end{equation}
and $\sigma_{xx}^{\mathcal{R}} = \sigma_{xx}^{\mathcal{I}} = 0$, 
$\sigma_{xy}^{\mathcal{I}} = 0$. 
In this low-frequency regime, the magneto-optical response of the
exchange field case becomes a special case of the quantizing magnetic
field case with $\nu_{\mathrm{T}} = \nu_{\mathrm{B}} = 1/2$ as explained above.

From Eqs.~(\ref{T_Efield})-(\ref{R_Efield}) we find the Faraday and Kerr angles
\begin{equation}
\theta_{\mathrm{F}} = \tan^{-1}\left[\left(\nu_{\mathrm{T}}+\nu_{\mathrm{B}}\right)\alpha\right],
\label{Fara0}
\end{equation}
\begin{equation}
\theta_{\mathrm{K}} = -\tan^{-1}\left[\frac{1}{\left(\nu_{\mathrm{T}}+\nu_{\mathrm{B}}\right)\alpha}\right].
\label{Kerr0}
\end{equation}
For total filling factor $\nu_{\mathrm{T}}+\nu_{\mathrm{B}}$ values that are not too large, the Faraday angle is quantized in integer multiples of the fine
structure constant
\begin{equation}
\theta_{\mathrm{F}} \simeq (\nu_{\mathrm{T}}+\nu_{\mathrm{B}})\alpha, \label{Fara_Quant}
\end{equation}
and the Kerr angle 
\begin{equation}
\theta_{\mathrm{K}} = -\frac{\pi}{2},
\label{piover2}
\end{equation}
becomes a full quarter polarization rotation.

It is also possible to understand Eq.~(\ref{piover2}) in terms of the
scattering mechanism of the reflected partial wave components. This 
understanding is crucial to see that Eq.~(\ref{piover2}) applies 
over a finite frequency range, as discussed in Section V B. 
The reflected electric field can be easily found from Eq.~(\ref{r20}). 
The algebra is simplified and the physics underlying Eq.~(\ref{piover2}) 
more easily illustrated when spatial-inversion symmetry across the TI film is preserved,
{\em i.e.} when top and bottom surfaces have the same 
conductivities. This happens when the exchange fields for both surfaces are
the same, or in the quantizing magnetic field case, when the surface
densities (and thus filling factors) are the same. Spatial-inversion symmetry then implies that
$\bar{r_{\mathrm{B}}} = \bar{r_{\mathrm{T}}}'$, $\bar{r_{\mathrm{T}}}
= \bar{r_{\mathrm{B}}}'$, $\bar{t_{\mathrm{B}}} =
\bar{t_{\mathrm{T}}}'$, and $\bar{t_{\mathrm{T}}} =
\bar{t_{\mathrm{B}}}'$. This allows the reflected 
electric field to be expressed solely in terms of the reflection and 
transmission matrix elements of one (e.g., the top) of the two surfaces. The first term on the right hand side of Eq.~(\ref{r20})
represents the partial wave directly reflected from the top surface, which we can evaluate in the low-frequency regime as 
\begin{eqnarray}
\tilde{E}^{\mathrm{r}0} =
\frac{1}{\left(1+\sqrt{\epsilon/\mu}\right)^2+\left(4\pi\sigma_{xy}^{\mathcal{R}}\right)^2}
\begin{bmatrix}
  1-\epsilon/\mu-\left(4\pi\sigma_{xy}^{\mathcal{R}}\right)^2 \\
  8\pi\sigma_{xy}^{\mathcal{R}}
\end{bmatrix}, \nonumber \\ 
\label{Er0}
\end{eqnarray}
where $\epsilon$, $\mu \simeq 1$ are the bulk dielectric constant and magnetic
permeability of the TI, and the second term constitutes all the partial waves that originate from successive reflections from the bottom surface  
\begin{eqnarray}
\tilde{E}^{\mathrm{r}'} &=&
  \frac{1}{1+\left(4\pi\sigma_{xy}^{\mathcal{R}}\right)^2} 
\begin{bmatrix}
 1 \\
  {4\pi\sigma_{xy}^{\mathcal{R}}}\end{bmatrix}
\label{Erp} \\
&&-\frac{1}{\left(1+\sqrt{\epsilon/\mu}\right)^2+\left(4\pi\sigma_{xy}^{\mathcal{R}}\right)^2}\begin{bmatrix} 
{2\left(1+\epsilon/\mu\right)} \\
{8\pi\sigma_{xy}^{\mathcal{R}}}
\end{bmatrix}. \nonumber
\end{eqnarray}
Summing the two contributions, we find that part of the second term on
the right-hand side of 
Eq.~(\ref{Erp}) cancels out the first term on the right-hand side of Eq.~(\ref{Er0}) completely,
yielding a total reflected field 
\begin{eqnarray}
\tilde{E}^{\mathrm{r}} =
\frac{1}{1+\left(4\pi\sigma_{xy}^{\mathcal{R}}\right)^2}\begin{bmatrix}
-\left(4\pi\sigma_{xy}^{\mathcal{R}}\right)^2 \\
4\pi\sigma_{xy}^{\mathcal{R}}
\end{bmatrix}.  \label{Ertotal}
\end{eqnarray}
Eq.~(\ref{Ertotal}) implies that the reflected partial waves that
originate from successive scattering off the bottom surface 
destructively interfere with the partial wave scattered off the top
surface, resulting in a suppression by a factor $\sim (\sigma_{xy}^{\mathcal{R}})^2$  of the reflected electric field
component along the incident polarization 
direction. This leads to the giant Kerr angle in Eq.~(\ref{piover2}). 
It is worthwhile to make clear that the large Kerr angle occurs because almost all of the reflected partial waves have a $90^{\circ}$ rotated polarization plane; it is not true, however, that almost all of the light is reflected.                

              
%
%


\subsection{Thick Film $d \gtrsim \lambda$}


In the previous section, we have focused on films with 
a thickness that is only a fraction of the wavelength. 
In this section, we shall relax this assumption and generalize our
considerations to thicker films, with thickness comparable to or
greater than the wavelength inside the film. 
Thick films do not in general show spectacular magneto-optical
effects because the Faraday and Kerr angles are suppressed by the
large dielectric constant of the TI bulk. Exceptions occur when the film thickness 
contains an integer multiple of half wavelengths inside the 
film, {\em i.e.} when the cavity resonance condition
$k_{\mathrm{TI}}d = N\pi$ is satisfied.  Here $k_{\mathrm{TI}} =
\sqrt{\epsilon\mu}\,\omega/c$ is the wave number in the TI film and $N
\ne 0$ is an integer. This property was first identified in Ref.~[\onlinecite{QiBulkSubstrate}], however 
the discussion there assumed an \textit{infinitely thick} dielectric
substrate underneath the TI film, neglecting scattering from the 
inevitable substrate-vacuum interface and thereby
overestimating substrate suppression of magneto-optical responses.
In this section, we first consider a free-standing thick film. We will 
then study the influence of a substrate, with its
finite thickness properly accounted for, in Section V. 

At resonance, a standing wave is established inside the film with the tangential components of the electric and magnetic
 fields on the interior of the top and bottom surfaces inside the film 
 related simply by a $\pm$ sign, {\em i.e.} $E_{\parallel}(z =
 -d/2^+)/E_{\parallel}(z = d/2^-) = (-1)^N$ 
%
%
(here $z = 0$ is taken at the center of the film), and similarly for $H$. Under such circumstances, 
the transmitted and reflected electric
fields become independent of the film's bulk dielectric
properties, and are found to be given by Eqs.~(\ref{T_Efield})-(\ref{R_Efield}) multiplied by a
phase factor $e^{-ik_0d}$, where $k_0 = \omega/c$ is the vacuum wave number. In contrast to the long-wavelength regime
we considered earlier in which the electromagnetic field varies 
slowly across the TI film, at resonance the field amplitudes 
change rapidly inside the film and the adiabatic condition $k_0d \ll
1$ does not apply. Regardless of the film thickness, however, the adiabaticity of TI surface electronic 
response can always be established at a sufficiently low frequency
satisfying $\omega_N \ll \Delta$ or $\Omega_{\mathrm{B}}$ 
[$\omega_N = N\pi c/(\sqrt{\epsilon\mu}\,d)$ is the cavity resonance frequency], such that the quantum Hall condition
Eq.~(\ref{Hall}) still holds. It follows from these considerations that the phases of the left-handed and right-handed circularly polarized
components of the transmitted and reflected light are given by 
%
%
%
\begin{eqnarray}
\mathrm{arg}(E_{\lambda}^{\mathrm{t}}) &=&
\frac{-{\lambda}\sin(k_0d)+2\pi\cos(k_0d)\sigma_{xy}^{\mathcal{R}}}{{\lambda}\cos(k_0d)+2\pi\sin(k_0d)
  \sigma_{xy}^{\mathcal{R}}}, \label{cr1} \\
\mathrm{arg}(E_{\lambda}^{\mathrm{r}}) &=&
\frac{\cos(k_0d)+{\lambda}2\pi\sin(k_0d) \sigma_{xy}^{\mathcal{R}}}{\sin(k_0d)-{\lambda}2\pi\cos(k_0d) \sigma_{xy}^{\mathcal{R}}},
%
\label{cr2}
\end{eqnarray}
where $\lambda = \pm 1$ labels the left and right-handed
circularly polarized light, and $\sigma_{xy}^{\mathcal{R}}$ contains the sum of the top 
and bottom surface Hall conductivities. 

Let us make several remarks here. If we set $k_0d \to 0$, Eqs.~(\ref{cr1})-(\ref{cr2}) coincide with the long-wavelength
($k_0d < k_{\mathrm{TI}}d \ll 1$) results, from which we recover 
Eqs.~(\ref{Fara0})-(\ref{Kerr0}) for the Faraday and Kerr angles. In
general, if in addition to the requirement $k_{\mathrm{TI}}d = N\pi$
for resonance we also have $k_0d = M\pi$, then Eqs.~(\ref{cr1})-(\ref{cr2})
would imply that $\theta_{\mathrm{F}} = \tan^{-1}\left[\left(\nu_{\mathrm{T}}+\nu_{\mathrm{B}}\right)\alpha\right]$
and $\theta_{\mathrm{K}} =
-\tan^{-1}\left[1/\left(\nu_{\mathrm{T}}+\nu_{\mathrm{B}}\right) \alpha\right]$
for integer $M$, and $\theta_{\mathrm{F}} =
-\tan^{-1}\left[1/\left(\nu_{\mathrm{T}}+\nu_{\mathrm{B}}\right) \alpha\right]$
and $\theta_{\mathrm{K}} =
\tan^{-1}\left[\left(\nu_{\mathrm{T}}+\nu_{\mathrm{B}}\right) \alpha\right]$
for half-odd integer $M$. These conditions would require
$\sqrt{\epsilon\mu}  = N/M$. With real materials this would seem to be
rather impossible, however with the advent of metamaterials 
it may be possible to engineer one with a matching dielectric constant
$(N/M)^2$, and employ it as an intervening dielectric between two
single-layer graphene sheets.  This point will be discussed further in Section
VI. In this light, we see that the long-wavelength limit $k_0d < k_{\mathrm{TI}}d \to
0$ is special because it automatically satisfies both conditions
$k_{\mathrm{TI}}d = N\pi$ and $k_0d = M\pi$ with $N = M = 0$. 

When $k_0d$ is not equal to a multiple of integer or half-odd integer
of $\pi$, which is generally the case, 
only the cavity resonance condition is satisfied and $k_0d$ cannot be assumed as small in Eqs.~(\ref{cr1})-(\ref{cr2}). 
Evaluating $\theta_{\mathrm{F}}$ and $\theta_{\mathrm{K}}$ 
from Eqs.~(\ref{FaraDef})-(\ref{KerrDef}) using Eqs.~(\ref{cr1})-(\ref{cr2}), 
we find that the Faraday and Kerr rotations at resonance have the same
universal quantized value
%
\begin{equation}
\theta_{\mathrm{F,K}} = \tan^{-1}\left[\left(\nu_{\mathrm{T}}+\nu_{\mathrm{B}}\right)\alpha\right].
\label{cr3}
\end{equation}
Note that Eqs.~(\ref{cr1})-(\ref{cr3}) do not depend on the value
of $N$ and therefore all cavity resonant modes yield the same Faraday and Kerr
rotations given by Eq.~(\ref{cr3}). It is important to emphasize, at
resonance, that although the Faraday angle is the same as the 
long-wavelength result Eq.~(\ref{Fara0}), the Kerr angle is not given
\cite{Remark1} by Eq.~(\ref{Kerr0}). Rather, both the Faraday and Kerr
rotations at resonance are given by the same quantized response in units of 
$\alpha$ like Eq.~(\ref{Fara0}). The giant Kerr effect
Eq.~(\ref{piover2}), therefore, is a unique long-wavelength
low-frequency property of the thin film system only. 

%
%



\section{Deviations from an Ideal Topological Insulator Film} 

Magneto-optical measurements of Faraday and 
Kerr rotations produced by topological insulators \cite{Drew,Armitage,Molenkamp} 
subjected to an external magnetic field have 
been performed by several groups.  The samples studied 
include bulk Bi$_2$Se$_3$ crystals \cite{Drew}, thin Bi$_2$Se$_3$
films \cite{Armitage} and strained HgTe films \cite{Molenkamp}.
It has not yet been possible to achieve ideal samples in which the 
half-quantized quantum Hall effect occurs, either in {\em DC} transport or 
optically.  In this section we consider non-ideal factors that often arise 
in experimental TI samples, and explain their consequences when they 
act independently.  We focus on the influence of bulk 
carriers, light scattering at oblique incidence, and the influence of a 
dielectric substrate. 

\subsection{Influence of Bulk Carrier Conduction}

Real TI samples are complicated by the presence of bulk free carriers
which are present because of unintentional doping by bulk defects 
\cite{Transport1,Transport2,Transport3,Transport4}.  
Recently some progress has been reported\cite{Oh1} in reducing the 
density of bulk carriers in TI thin films. 

Bulk conduction can be described by
a complex bulk dielectric function $\epsilon(\omega)$ which is
related to the bulk conductivity $\Sigma(\omega)$ by
\begin{equation}
\epsilon(\omega) =
\epsilon_{\mathrm{b}}+i\frac{4\pi}{\omega}\Sigma(\omega),
\label{bulkepsilon}
\end{equation}
where $\epsilon_{\mathrm{b}}$ is the high-frequency dielectric
constant of the TI.
When the quantized Hall regime is approached on the TI surfaces,
the bulk Hall angle $\tan^{-1}(\Sigma_{xy}/\Sigma_{xx})$ is expected to be 
much smaller than the surface Hall angle (which
becomes infinite when $\sigma_{xx} \to 0$).
For definiteness the numerical results reported below assume 
that the longitudinal bulk conductivity dominates, 
and that its frequency-dependence can be described by the Drude-Lorentz form,
\begin{equation}
\Sigma(\omega) =
\frac{\Omega_{\mathrm{b}}^2}{4\pi\left(1/\tau_{\mathrm{b}}-i\omega\right)},
\label{bulkcond}
\end{equation}
where $\Omega_{\mathrm{b}} = \sqrt{{4\pi N_{\mathrm{b}}e^2}/{m_{\mathrm{b}}}}$ is the plasma frequency of
the bulk carriers (with density $N_{\mathrm{b}}$ and effective mass $m_{\mathrm{b}}$), and $1/\tau_{\mathrm{b}}$ is the disorder
scattering rate due to impurities present in the bulk.  

The influence of a finite bulk conductivity is particularly simple to describe in the 
long-wavelength low-frequency limit of Eqs.~(\ref{T_Efield})-(\ref{R_Efield}).  
The total current integrated across the TI film in
Eq.~(\ref{LongWave}) now picks up an extra bulk conductivity
contribution given by $\Sigma d$ ($d$ is the film
thickness), in addition to the conductivities from the two surfaces.  The 
change in the expressions for the transmitted and reflected electric fields 
Eqs.~(\ref{T_Efield})-(\ref{R_Efield}) is therefore altered by the replacement
$\sigma_{xx} \to \sigma_{xx}+\Sigma d/c$. 
When the surface has a perfect quantum Hall effect, the 
modified expressions for the Faraday and Kerr
angles are 
\begin{eqnarray}
\theta_{\mathrm{F}} &=&
\tan^{-1}\left[\frac{\left(\nu_{\mathrm{T}}+\nu_{\mathrm{B}}\right)\alpha}{1+2\pi\Sigma
d/c}\right], \label{Faraw0_bulk}
\\
\theta_{\mathrm{K}} &=&
\tan^{-1}\left\{\frac{4\left(\nu_{\mathrm{T}}+\nu_{\mathrm{B}}\right)\alpha}{1-(1+4\pi\Sigma
  d/c)^2-[2\left(\nu_{\mathrm{T}}+\nu_{\mathrm{B}}\right)\alpha]^2}\right\},
\nonumber \\
\label{Kerrw0_bulk}
\end{eqnarray}
where $\Sigma = \Sigma(0)$ is the bulk \textit{DC} conductivity. The 
bulk carriers thus enter as an \textit{effective} longitudinal surface conductivity 
$\Sigma d$. Eq.~(\ref{Faraw0_bulk}) implies that 
the influence of bulk conduction is negligible on the Faraday effect
when 
\begin{equation}
\frac{\Sigma d}{(e^2/h)} \ll 1/\alpha. \label{Faracriterion} 
\end{equation}
For the Kerr effect, Eq.~(\ref{Kerrw0_bulk})  implies a stricter condition 
for negligible bulk conduction:
\begin{equation}
\frac{\Sigma d}{(e^2/h)} \lesssim \alpha. \label{Kerrcriterion}
\end{equation}
When the bulk conductivity is sufficiently small that
Eq.~(\ref{Kerrcriterion}) is satisfied,
Eqs.~(\ref{Faraw0_bulk})-(\ref{Kerrw0_bulk}) reduce to 
the universal results for the Faraday and Kerr effects,
Eqs.~(\ref{Fara0})-(\ref{piover2}). 

\begin{figure}
  \includegraphics[width=8.5cm,angle=0]{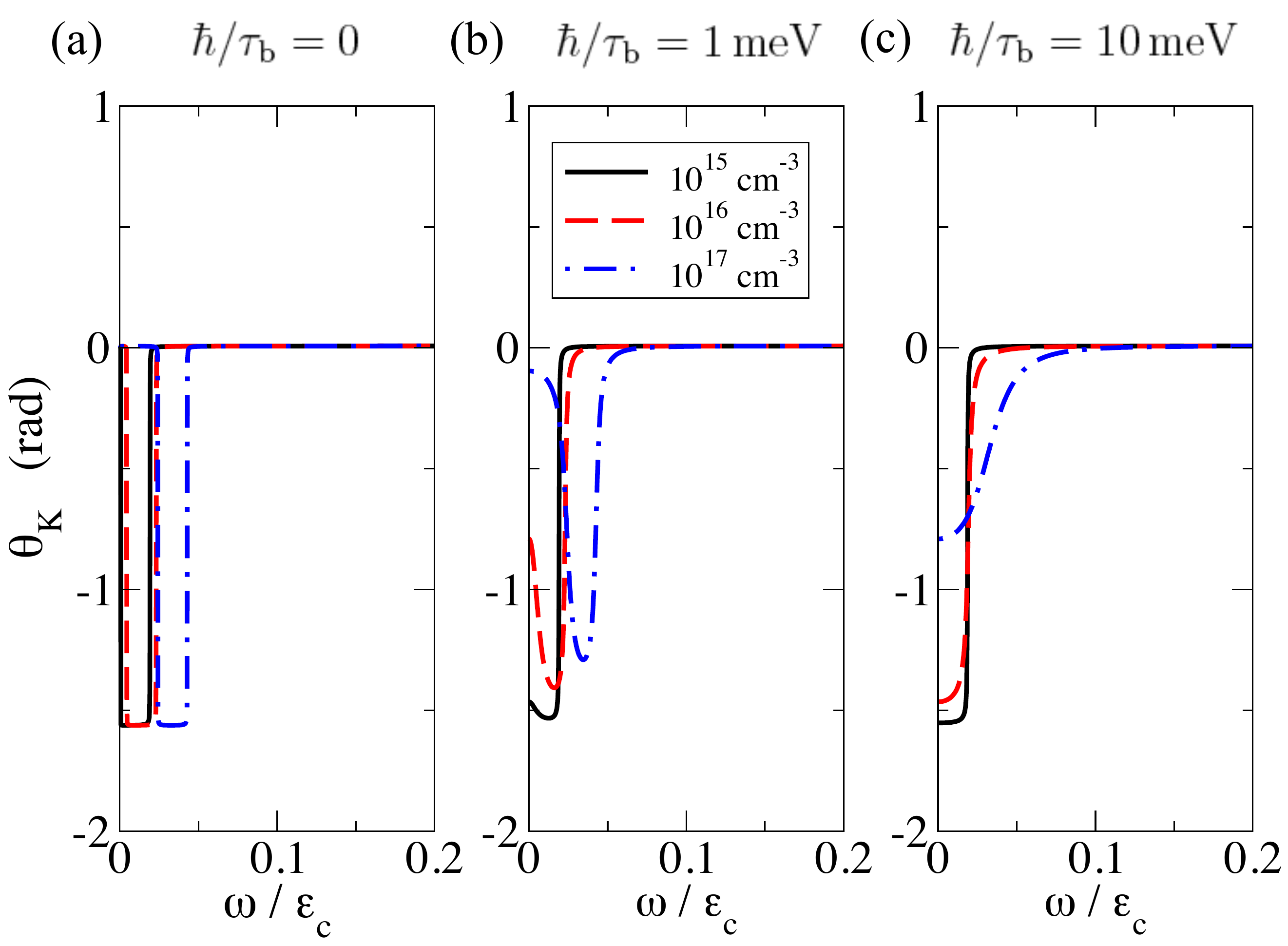}
\caption{(Color online) Kerr rotation for top and bottom surface densities $n_{\mathrm{T}} = n_{\mathrm{B}} =
  3\times10^{11}\,\mathrm{cm}^{-2}$ and filling factors $\nu_{\mathrm{T}} = \nu_{\mathrm{B}}
  = 1/2$ (corresponding to a magnetic field of $25\,\mathrm{T}$) for (a) no bulk carrier scattering
  (b) $\hbar/\tau_{\mathrm{b}} = 1\,\mathrm{meV}$, and (c) $10\,\mathrm{meV}$ at different values of 
  bulk carrier densities $N_{\mathrm{b}} = 10^{15}\,\mathrm{cm}^{-3} (\textrm{black
    solid line})$, $10^{16}\,\mathrm{cm}^{-3} (\textrm{red dashed})$,
  $10^{17}\,\mathrm{cm}^{-3} (\textrm{blue dot-dashed})$. 
The dielectric constant of Bi$_2$Se$_3$ $\epsilon_{\mathrm{b}} = 29$,
 bulk carrier effective mass $m_{\mathrm{b}} = 0.15m_e$ \cite{Transport4,Eto,Butch}, film
 thickness $d = 30\,\mathrm{nm}$.}
\label{fig_Kerr_Bulk}
\end{figure}

Eq.~(\ref{Kerrcriterion}) can 
alternately can be expressed as a condition that has to be satisfied by
the bulk carrier density: 
\begin{equation}
N_{\mathrm{b}} \lesssim \frac{\alpha m_{\mathrm{b}}}{h
  \tau_{\mathrm{b}} d}. \label{Nb}
\end{equation}
For a $30$-nm thick Be$_2$Se$_3$ film and disorder broadening
$\hbar/\tau_{\mathrm{b}} = 1-10\,\mathrm{meV}$, we estimate that the
to observe the giant Kerr effect, the bulk carrier
density 
must be smaller than $10^{14}\,-\,10^{15}\,\mathrm{cm}^{-3}$. 
To reach the regime of the quantized Faraday effect
given by Eq.~(\ref{Faracriterion}), the bulk carrier density is allowed to be larger by a factor $1/\alpha^2$ so that  
$N_{\mathrm{b}} \lesssim {m_{\mathrm{b}}}/({\alpha h
  \tau_{\mathrm{b}} d}) \simeq 
10^{18}\,-\,10^{19}\,\mathrm{cm}^{-3}$. Fig~\ref{fig_Kerr_Bulk} shows 
the low-frequency Kerr angle in the presence of bulk
conduction. The magneto-optical response is modified by the presence
of bulk carriers principally in the low frequency regime where the bulk Drude-Lorentz conductivity
is peaked.

Because bulk carriers originate
from bulk defects, $N_{\mathrm{b}}$ and
$\tau_{\mathrm{b}}$ are related.  For the purpose of studying their
influence on the magneto-optical response we will nevertheless treat $N_{\mathrm{b}}$ and
$\tau_{\mathrm{b}}$ as independent parameters. We first illustrate the
case when there are no impurities in the bulk, {\em i.e.} the case of a bulk free plasma.  
We find that the giant Kerr angle remains but undergoes a shift to
progressively higher frequencies for increasing bulk carrier density
[Fig.~\ref{fig_Kerr_Bulk} (a)].    
Including bulk impurities broadens 
[Figs.~\ref{fig_Kerr_Bulk} (b)-(c)] the giant Kerr
effect, but the Kerr angle remains
substantial $\sim 1\,\mathrm{rad}$ for a bulk density $N_{\mathrm{b}}
= 10^{17}\,\mathrm{cm}^{-3}$ and disorder broadening
$\hbar/\tau_{\mathrm{b}} = 1-10\,\mathrm{meV}$.
%

%
%

%


For thick films, we note that the presence of bulk conduction the cavity
resonance condition becomes non-trivial.  For this reason there 
is no simple analytic criterion for neglecting the bulk conductivity.
Experimentally, this may also present a challenge 
since resonance frequencies cannot be readily estimated unless the bulk
conductivity is known from a separate transport measurement.

In passing, we mention that bulk optical phonon modes of the
topological insulator can also be excited at higher frequencies. Since
phonon energies are specific to different TI materials and we are mainly interested in the low-frequency
regime for the magneto-optical effects, we shall neglect the bulk
optical phonon contributions to the conductivity.
Phonon effects can
be modeled 
by including an additional term \cite{Drew,Armitage}
%
\begin{equation}
\epsilon_{\mathrm{ph}}(\omega) = \frac{f_{\mathrm{ph}}}{{\omega_{\mathrm{ph}}}^2-\omega^2-i\omega/2\tau_{\mathrm{ph}}},
\label{bulkphonon}
\end{equation}
to the dielectric constant Eq.~(\ref{bulkepsilon}), where
$f_{\mathrm{ph}}$ is the spectral weight of the phonon mode with
frequency $\omega_{\mathrm{ph}}$, and $1/2\tau_{\mathrm{ph}}$
is the phonon damping rate. This separation between the 
electronic and the phononic contributions to the bulk dielectric
function 
applies only when the electronic time
scales ($\Omega_{\mathrm{b}}^{-1}, \varepsilon_{F\mathrm{b}}^{-1}$, where
$\varepsilon_{F\mathrm{b}}$ is the bulk Fermi energy) are 
much smaller than the phonon time scale ($\omega_{\mathrm{ph}}^{-1}$),
making plasmon-phonon coupling negligible. 



\subsection{Influence of Oblique Incidence and the Substrate}


%
%
In this section, we examine the effects of oblique incidence and of the 
substrate on the magneto-optical effects. This discussion is particularly germane to
experiments because oblique incidence may afford an 
advantage over normal incidence for Kerr effect measurement as it allows for spatial separation 
between the incident light source and the reflected light polarizer,
and enhances the reflected light intensity.


\subsubsection{Thin TI Film and Thin Substrate}
We now account for a dielectric substrate layer underneath the TI film and 
allow for the (semi-infinite) medium underneath the substrate to be
different from vacuum, solely for generality. 
The incident angle on the TI film and the emerging angle from the
substrate can therefore assume different values, denoted by
$\theta_{\mathrm{i}}$ and $\theta_{\mathrm{o}}$ respectively. 
At low frequencies ($\omega \ll \Delta$ or $\Omega_B$) and long wavelength 
compared to both the film thickness $d$ and substrate thickness
$d_{\mathrm{s}}$, the analysis is again simple. 
We find the following Faraday and
Kerr angles under oblique incidence 
(Expressions for the transmission and reflection coefficients at
oblique incidence are presented in the Appendix.):
\begin{equation}
\theta_{\mathrm{F}} =
\mathrm{tan}^{-1}\left[\frac{2\alpha(\nu_{\mathrm{T}}+\nu_{\mathrm{B}})\mathrm{cos}\theta_{\mathrm{i}}}{\mathrm{cos}\theta_{\mathrm{i}}+\mathrm{cos}\theta_{\mathrm{o}}}\right], \label{Faraw0_ob} 
\end{equation}
\begin{eqnarray}
&&\theta_{\mathrm{K}} = \label{Kerrw0_ob} \\
&&-\mathrm{tan}^{-1}\left[\frac{8\alpha(\nu_{\mathrm{T}}+\nu_{\mathrm{B}})\mathrm{cos}\theta_{\mathrm{i}}\mathrm{cos}\theta_{\mathrm{o}}}{\mathrm{cos}2\theta_{\mathrm{o}}-\mathrm{cos}2\theta_{\mathrm{i}}+8\alpha^2(\nu_{\mathrm{T}}+\nu_{\mathrm{B}})^2\mathrm{cos}\theta_{\mathrm{i}}\mathrm{cos}\theta_{\mathrm{o}}}\right],
\nonumber
\end{eqnarray}
%
It is important to recognize that Eqs.~(\ref{Faraw0_ob})-(\ref{Kerrw0_ob}) are 
independent of the bulk dielectric constants of not only the TI
film, but also importantly the substrate. 
The transmitted and reflected fields are generally dependent on the dielectric
constant of the ambient medium surrounding the TI film however; these 
dependences enter the Faraday and Kerr angles expressions through the incident and emergent angles
$\theta_{\mathrm{i}}$ and $\theta_{\mathrm{o}}$. 
Since the measurement apparatus is almost inevitably 
located in vacuum, so one has $\theta_{\mathrm{o}} =
\theta_{\mathrm{i}}$ by Snell's law.  It is easy to verify that 
Eqs.~(\ref{Faraw0_ob})-(\ref{Kerrw0_ob}) then
reduce to the universal results in Section IV A, 
and it follows that the long-wavelength low-energy results are not influenced by the angle of 
incidence or the presence of a thin ($d_{\mathrm{s}} \ll \lambda$) dielectric substrate. 

The giant Kerr effect survives \cite{Tse_1,Tse_2} up to a relatively large frequency 
which we refer to as the Kerr frequency $\omega_{\mathrm{K}}$. First we derive an analytic formula for
$\omega_{\mathrm{K}}$ at normal incidence that also allows for the
presence of a substrate. For small frequencies, the reflected circularly 
polarized components can be decomposed into separate
leading-order contributions from the top and bottom TI surfaces, the
bulk TI dielectric, and the substrate dielectric:
\begin{eqnarray}
E_{\pm}^{\mathrm{r}} &\simeq& 
\frac{i\omega}{2c}\left[\left(\epsilon-\mu\right)d+\left(\epsilon_{\mathrm{s}}-\mu_{\mathrm{s}}\right)d_{\mathrm{s}}\right]
-i2\pi{\sigma_{xx}^{\mathcal{I}}}'(0)\omega \nonumber \\
&&\pm i2\pi\sigma_{xy}^{\mathcal{R}}(0),
\label{freqwin1}
\end{eqnarray}
where $\sigma_{xx},\sigma_{xy}$ contain the top and bottom TI surface conductivities, 
$\epsilon_{\mathrm{s}}, \mu_{\mathrm{s}}$ are the dielectric constant
and permeability ($= 1$) of the substrate, and $'$ in 
$\sigma_{xx}^{\mathcal{I}}$ denotes a frequency derivative. 
The real components of $E_{\pm}^{\mathrm{r}}$ are smaller by a factor $\sim\alpha$ in this
regime. As frequency increases, the dielectric contribution
to the imaginary part of $E_{\pm}^{\mathrm{r}}$ eventually dominates so that
the $\pm$ components have the same sign and $\theta_{\mathrm{K}}$ rapidly falls
to a small value. The frequency range for which giant Kerr
angles occur is approximately given by
\begin{equation}
\omega_{\mathrm{K}} = 
\frac{2\pi\sigma_{xy}^{\mathcal{R}}(0)}{\left[\left(\epsilon-\mu\right)d+\left(\epsilon_{\mathrm{s}}-\mu_{\mathrm{s}}\right)d_{\mathrm{s}}\right]/2c-2\pi{\sigma_{xx}^{\mathcal{I}}}'(0)}. \label{freqwin_sub}
\end{equation}
%
%
A similar analysis can be carried out for the case of oblique
incidence. 
The Kerr frequency $\theta_{\mathrm{K}}$ at oblique incidence 
without substrate is given by  
\begin{eqnarray}
\omega_{\mathrm{K}} = 
\frac{2\pi\sigma_{xy}^{\mathcal{R}}(0)\cos\theta_{\mathrm{i}}}{\left(\epsilon\cos^2\theta-\mu\cos^2\theta_{\mathrm{i}}\right)d/2c-2\pi{\sigma_{xx}^{\mathcal{I}}}'(0)}. \label{freqwin_ob} 
\end{eqnarray}
where $\theta_{\mathrm{i}}$ and $\theta$ are the incident angle and refracted
angle inside the TI film, respectively, related by Snell's law 
$\sin\theta_{\mathrm{i}} = \sqrt{\epsilon\mu}\sin\theta$. 

\begin{figure}
  \includegraphics[width=8.5cm,angle=0]{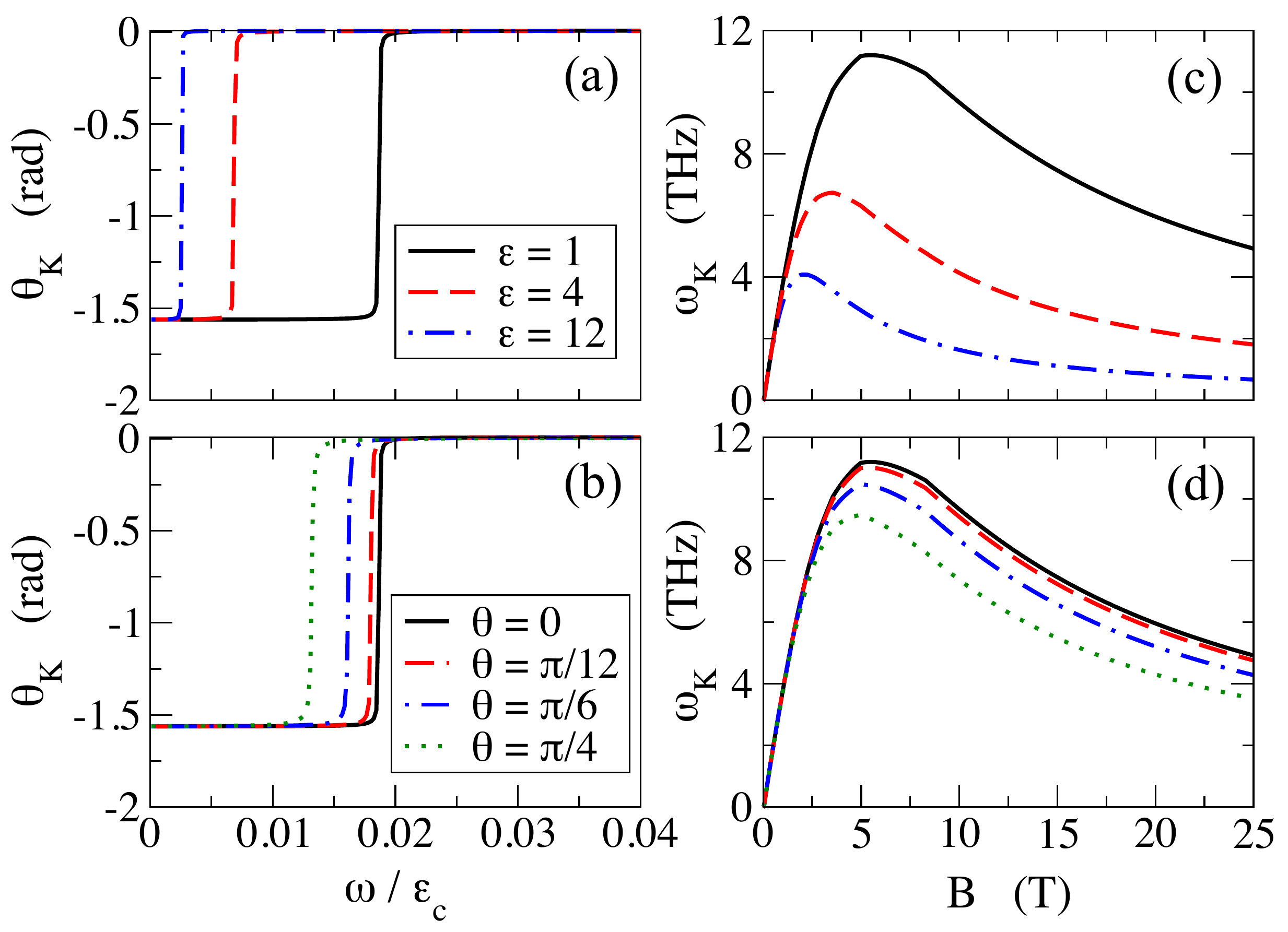}
\caption{(Color online) Effect of substrate and oblique incidence on
  the low-frequency Kerr rotation. (a)-(b) illustrate the Kerr angle versus
  frequency for two cases: (a) normal incidence on a TI film with a
  dielectric substrate of thickness $d_{\mathrm{s}} = 0.5\,\mu\mathrm{m}$: $\epsilon = 1$
  (a free-standing sample), $\epsilon = 4$ (a SiO$_2$ substrate), $\epsilon = 12$
  (a Si substrate), and (b) oblique incidence on a free-standing TI film at
  incidence angle $\theta_i = 0, \pi/12, \pi/6, \pi/4$. 
The top and surface densities are $n_{\mathrm{T}} = n_{\mathrm{B}} =
3\times10^{11}\,\mathrm{cm}^{-2}$ and $B = 25\,\mathrm{T}$. 
 (c)-(d) show respectively the Kerr frequency as a function of magnetic field corresponding to
  cases (a) and (b) at the same surface densities. The TI film
  thickness $d = 30\,\mathrm{nm}$.}
\label{fig_Kerr_Subs_Oblq}
\end{figure}

Eqs.~(\ref{freqwin_sub})-(\ref{freqwin_ob}) show that oblique
incidence and the presence of a substrate reduce the frequency window
over which the giant Kerr angles occur. This is illustrated numerically in 
Fig.~(\ref{fig_Kerr_Subs_Oblq}) where we have calculated the Kerr angle as a function of frequency and magnetic field for different
dielectric substrates and different values of incidence angle. 

\subsubsection{Thin TI film and Thick Substrate}
Above we considered the case when the substrate thickness is
small compared with the wavelength. We see that as long as the
substrate thickness remains smaller than the wavelength, increasing
the thickness only suppresses the Kerr frequency window, but the $\pi/2$ rotation
at very small frequencies survives.  Experimentally, however, one may
have to employ a substrate with supra-wavelength thickness for
various reasons; this motivates us to 
consider the effect of a thicker substrate. When the substrate 
thickness is increased beyond one wavelength, one can
expect that the magnitude of the giant Kerr rotation is
suppressed.  But that is not the end of the story. 
Indeed, a logic similar to that employed in Section IV B tells us that when the substrate thickness contains
an integer number of half wavelength, the 
resulting Faraday and Kerr rotations will again be independent of the
substrate dielectric properties. 

For wavelength short compared with the
substrate thickness, but still long compared with the TI film thickness
($k_{\mathrm{TI}}d \ll 1$) and $\omega \ll \Delta$ or $\Omega_{\mathrm{B}}$,
we find the following phases for the left- and right-handed ($\lambda 
= \pm$) circularly polarized transmitted and reflected light for normal incidence 
\begin{widetext}
\begin{eqnarray}
&&\mathrm{arg}\left(E _{\lambda}^{\mathrm{t}}\right) = \nonumber \\
&&\tan^{-1}\frac{2 Z_{\mathrm{s}}^{-1}\cos(k_{\mathrm{s}}d_{\mathrm{s}})\left[\lambda
      2\pi\sigma_{xy}^{\mathcal{R}}\cos(k_0d_{\mathrm{s}})-\sin(k_0d_{\mathrm{s}})\right]
+\sin(k_{\mathrm{s}}d_{\mathrm{s}})\left[\left(1+Z_{\mathrm{s}}^{-2}\right)\cos(k_0d_{\mathrm{s}})+\lambda
      4\pi\sigma_{xy}^{\mathcal{R}}\sin(k_0d_{\mathrm{s}})\right]}
{2 Z_{\mathrm{s}}^{-1}\cos(k_{\mathrm{s}}d_{\mathrm{s}})\left[\lambda
      2\pi\sigma_{xy}^{\mathcal{R}}\sin(k_0d_{\mathrm{s}})+\cos(k_0d_{\mathrm{s}})\right]
+\sin(k_{\mathrm{s}}d_{\mathrm{s}})\left[\left(1+Z_{\mathrm{s}}^{-2}\right)\sin(k_0d_{\mathrm{s}})-\lambda
      4\pi\sigma_{xy}^{\mathcal{R}}\cos(k_0d_{\mathrm{s}})\right]}, \label{ThickSubs1} \\
&&\mathrm{arg}\left(E _{\lambda}^{\mathrm{r}}\right) =
-\tan^{-1}\left\{ \right. \label{ThickSubs2} \\ 
&&\left.\frac{2\left\{\left(Z_{\mathrm{s}}^{-2}+1\right)\lambda 4\pi\sigma_{xy}^{\mathcal{R}}+\left(Z_{\mathrm{s}}^{-2}-1\right)\left[
      \lambda 4\pi\sigma_{xy}^{\mathcal{R}}\cos(2k_{\mathrm{s}}d_{\mathrm{s}})+Z_{\mathrm{s}}^{-1}\sin(2k_{\mathrm{s}}d_{\mathrm{s}})\right]\right\}}
{\left(Z_{\mathrm{s}}^{-2}+1\right)[Z_{\mathrm{s}}^{-2}-1+\left(4\pi\sigma_{xy}^{\mathcal{R}}\right)^2]-\left(Z_{\mathrm{s}}^{-2}-1\right)[Z_{\mathrm{s}}^{-2}+1-\left(4\pi\sigma_{xy}^{\mathcal{R}}\right)^2]\cos(2k_{\mathrm{s}}d_{\mathrm{s}})
+Z_{\mathrm{s}}^{-1}\left(Z_{\mathrm{s}}^{-2}-1\right) \lambda 8\pi\sigma_{xy}^{\mathcal{R}}\sin(2k_{\mathrm{s}}d_{\mathrm{s}})}\right\}, \nonumber
\end{eqnarray}
\end{widetext}
where for notational simplicity we have defined the wave impedance
$Z_{\mathrm{s}} = \sqrt{\mu_{\mathrm{s}}/\epsilon_{\mathrm{s}}}$ for
the substrate. 
\begin{figure}
  \includegraphics[width=8.5cm,angle=0]{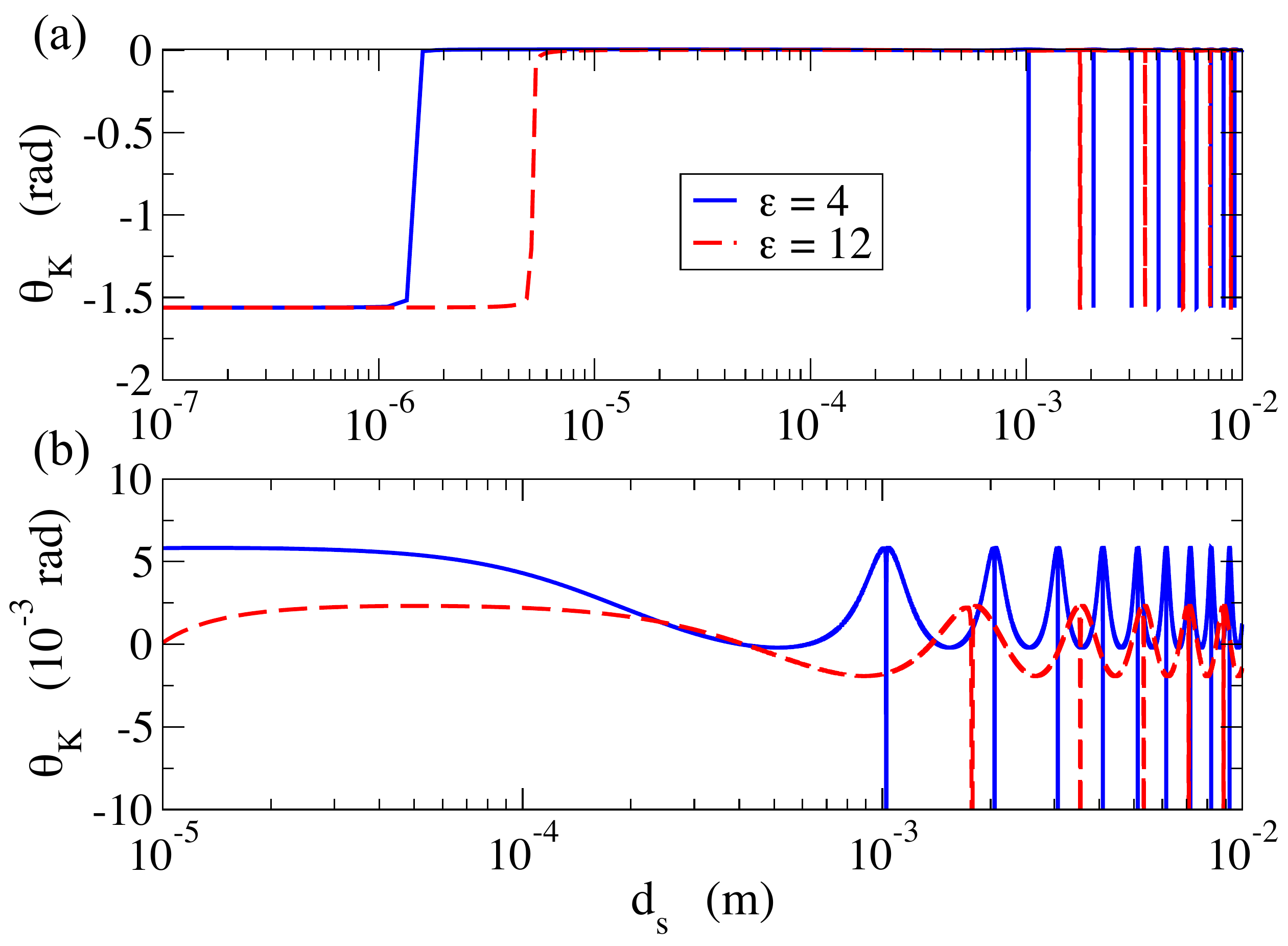}
\caption{(Color online) (a). Dependence of the low-frequency Kerr rotation of a TI thin film 
   on the substrate thickness $d_{\mathrm{s}}$
   for frequency $\omega/\varepsilon_{\mathrm{c}} = 10^{-3}$, and 
   substrate dielectric constants $\epsilon = 4$ (SiO$_2$), $\epsilon = 12$
  (Si). At this frequency, $\lambda \simeq 7\,\mathrm{mm}$ is always
  much longer than the TI film thickness, but becomes comparable to the
  substrate thickness when $d_{\mathrm{s}} \sim
  \lambda/\sqrt{\epsilon_{\mathrm{s}}\mu_{\mathrm{s}}}$. Cavity
  resonances of the giant Kerr rotation can be seen at values of
  $d_{\mathrm{s}}$ equal to integer multiples of
  $\lambda/2\sqrt{\epsilon_{\mathrm{s}}\mu_{\mathrm{s}}}$. 
(b). Close-up
  showing the finer features of $\theta_{\mathrm{K}}$ additional to the
  cavity resonances; Fabry-Perot type oscillations are clearly
  seen. For thick substrates, the value of $\theta_{\mathrm{K}}$ is strongly
  suppressed compared to the long-wavelength result away from 
  the cavity resonance values. 
The values of $n_{\mathrm{T}}$, $n_{\mathrm{B}}$, and $d$ are
  the same as in Fig.~\ref{fig_Kerr_Subs_Oblq}.} 
\label{fig_Kerr_Subs_ds}
\end{figure}
If, in addition, we impose the 
requirement that the substrate thickness contains an integer multiple of
half wavelengths ($k_{\mathrm{s}}d_{\mathrm{s}} = N\pi$, $N \ne 0$)   
then Eqs.~(\ref{ThickSubs1})-(\ref{ThickSubs2}) greatly simplify, 
yielding 
\begin{eqnarray}
\mathrm{arg}\left(E _{\lambda}^{\mathrm{t}}\right) &=&
\tan^{-1}\frac{\lambda
  2\pi\sigma_{xy}^{\mathcal{R}}\cos(k_0d_{\mathrm{s}})-\sin(k_0d_{\mathrm{s}})}{\lambda
  2\pi\sigma_{xy}^{\mathcal{R}}\sin(k_0d_{\mathrm{s}})+\cos(k_0d_{\mathrm{s}})}, \label{ThickSubs3} \\
\mathrm{arg}\left(E _{\lambda}^{\mathrm{r}}\right) &=&
-\lambda\tan^{-1}\frac{1}{2\pi\sigma_{xy}^{\mathcal{R}}},
\label{ThickSubs4}
\end{eqnarray}
from which we recover the quantized Faraday [Eq.~(\ref{Fara0})] and giant Kerr rotations [Eq.~(\ref{Kerr0})]. This
tells us that the Faraday and Kerr rotations can survive
suppression effects from a thick substrate as long as the
substrate thickness satisfies the resonance condition. 
Although it is remarkable that Eqs.~(\ref{Fara0})-(\ref{Kerr0}) still hold in 
this circumstance, it is important that the light frequency needs to
be precisely tuned to the resonant frequency of the substrate.
In contrast, if one has the liberty to use a thin substrate, the giant
Kerr rotation can be observed in a relatively broad range of
frequencies up to $\omega_{\mathrm{K}}$ [Eq.~(\ref{freqwin_sub})]. 
Fig.~\ref{fig_Kerr_Subs_ds} shows the Kerr angle calculated as a
function of substrate thickness. We see that the Kerr angle remains $\pi/2$ for substrate
thickness $d_{\mathrm{s}}$ much smaller than the wavelength (up to $\sim
1\,\mu\mathrm{m}$ in the plot) and then becomes suppressed for larger
substrate thickness. However, when $d_{\mathrm{s}}$ becomes comparable
to the wavelength in the substrate, a series of sharply-defined cavity
resonance peaks is seen that preserves the giant $\pi/2$ value
[Fig.~\ref{fig_Kerr_Subs_ds} (a)]. Around the same
range of $d_{\mathrm{s}}$ values, Fabry-perot like oscillations of the
Kerr rotation are also clearly seen in Fig.~\ref{fig_Kerr_Subs_ds} (b). 

%
%


\section{Generalization to Other Systems with Quantized Hall Conducting Layers}

%
\begin{figure}
  \includegraphics[width=6.5cm,angle=0]{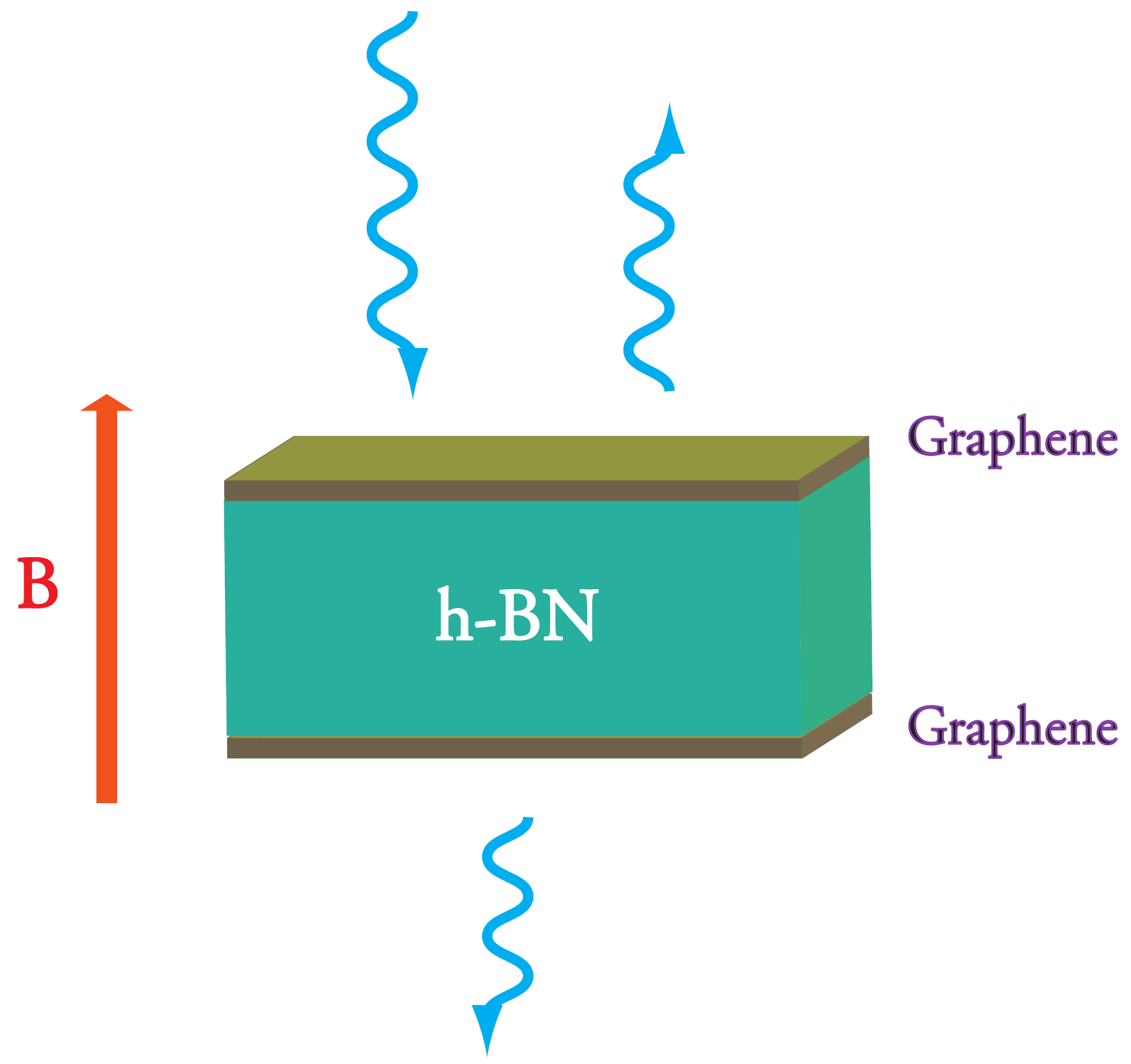}
\caption{(Color online) Experimental setup of the double-layer
  graphene heterostructure sandwiching a hexagonal boron nitride
  (h-BN) substrate. A quantizing magnetic field is applied
  perpendicularly to the layers.}
\label{Graphene}
\end{figure}

Because topological insulator properties are at present still obscured by the issue of bulk
conduction \cite{Transport1,Transport2,Transport3,Transport4}, and because samples do not 
yet have the quality necessary to yield strongly developed quantum Hall effects,
it is natural to ask if similar magneto-optical effects can be achieved in other materials systems.
Indeed, the magneto-optical effects we have discussed are not essentially 
distinct from those of other systems with two nearby conducting layers that exhibit quantum 
Hall effects.  Eqs.~(\ref{Fara0}), (\ref{Kerr0}), (\ref{cr3}) for the Faraday and
Kerr rotations in Section IV apply to a wide variety of other systems when they are 
placed in an external magnetic field. 

 
Systems with two nearby graphene layers appear to be particularly attractive 
because they are also described by massless Dirac equations and, like TI surface states,
quite sensitive to time-reversal symmetry breaking perturbations.  In fact,
the quantum Hall effect can be realized in graphene sheets at magnetic field 
strengths that are so low \cite{QH_Kim1,QH_Kim2,Geim1} 
that applications in optics are not out of the question.  
Aside from integer and fractional quantum Hall effects in 
external magnetic fields, monolayer and bilayer graphene can also 
potentially exhibit quantized anomalous Hall effects\cite{Graphene_QAHE1,Graphene_QAHE2,Franz} 
due to surface adsorption of transition metal atoms. 

One experimental system that has now been realized experimentally \cite{Kim_BN}, but not 
yet studied optically, consists of 
two graphene layers separated by a few layers of hexagonal boron
nitride.
Transport experiments have demonstrated that the quantum Hall effect is 
already strong in this type of system at fields well below $1$
Tesla. We propose the experimental setup shown in Fig.~\ref{Graphene} 
to observe the dramatic magneto-optical effects. 
Because of valley and approximate spin degeneracies the strongest 
quantum Hall effects occur at filling factor $\nu_{\mathrm{T,B}}=\pm 2$, rather 
than at $\nu_{\mathrm{T,B}}=\pm 1/2$, but this only changes some details of the 
magneto-optical properties.  Indeed, the change in filling factors might 
be desirable for some potential applications.  Realizing systems with two
(or indeed many) essentially decoupled layers separated by much less than
a wavelength is feasible.  The small spacing between 
essentially isolated quantum Hall layers increases
the frequency window over which strong magneto-optical effects are anticipated. 
In addition bulk conduction is automatically eliminated. 

\section{Conclusion}

We have presented a comprehensive theory for the magneto-optical
Faraday and Kerr effects of topological insulator films, and more generally
of layered quantized Hall systems. We identify a {\em topological} 
regime in which the light frequency is low compared to surface gaps 
opened up by time-reversal symmetry breaking perturbations and the 
light wavelength is either long compared to the film thickness or 
an integer multiple of twice the film thickness.
In the topological regime, the magneto-optical effects are
dramatic and universal. For thin films, the Faraday rotation angle is 
quantized in units of the fine structure constant, and the Kerr
angle exhibits a giant $90$ degrees rotation. For thick films that
contain a commensurate number of half wavelength, both the Faraday and
Kerr rotations are quantized in units of the fine structure
constant. In the presence of bulk conduction, the dramatic Faraday and
Kerr effects for thin films remain robust as long as the effective
two-dimensional conductivity from the bulk, in $e^2/h$ units, is smaller than the fine
structure constant. The effect of a thick substrate, which may sometimes 
be experimentally necessary, can be nullified either by making it thinner 
than a light wavelength or, if it must be thick, by tuning its thickness to 
an integer number of half wavelength. The giant Kerr effect 
remains unaffected by oblique incidence when a thin film with a thin 
substrate is used. These magneto-optical effects can also be realized, 
perhaps even more readily, in systems with two graphene layers
separated by hexagonal boron nitride 
or another thin dielectric.

\section{Acknowledgement}

This work was supported by the Welch under Grant No.
F1473 and by DOE under Grant No. DE-FG03-02ER45985.
We are indebted to many for their useful discussions throughout the progress of
this work: N. Peter Armitage, Kenneth Burch, Dennis Drew, Jim Erskine, 
Zhong Fang, Jun Kono, Joel Moore, Xiao-Liang Qi, Gennady Shvets, Rolando Vald\'es Aguilar, David
Vanderbilt, and Shou-Cheng Zhang.

\newpage
\section{Appendix. Transmission and Reflection Coefficients at Oblique
Incidence}

We denote the incident and emergent angles of the electromagnetic wave 
after scattering with the interface by $\theta_i$ and $\theta_j$. The 
matrix elements of the reflection and transmission tensors can be
found as 
%
%
\begin{widetext}
\begin{eqnarray}
r_{xx} &=& \frac{e^{i2k_ia_i\cos\theta_i}}{D_{ji}} \left\{\sqrt{\frac{\epsilon_i}{\mu_i}}\cos^2\theta_i\sec\theta_j\left(4\pi{\sigma}_{xx}+\sqrt{\frac{\epsilon_j}{\mu_j}}\sec\theta_j\right)-\sqrt{\frac{\epsilon_i}{\mu_i}}\left(\sqrt{\frac{\epsilon_j}{\mu_j}}+4\pi{\sigma}_{xx}\sec\theta_j\right)\right.
\nonumber \\
&&\left.-\cos\theta_i\left\{4\pi\sqrt{\frac{\epsilon_j}{\mu_j}}{\sigma}_{xx}+\sec\theta_j\left[-\frac{\epsilon_i}{\mu_i}+\frac{\epsilon_j}{\mu_j}+16\pi^2\left({\sigma}_{xx}^2+{\sigma}_{xy}^2\right)+4\pi\sqrt{\frac{\epsilon_j}{\mu_j}}{\sigma}_{xx}\sec\theta_j\right]\right\}\right\},
\nonumber \\
r_{yy} &=& -\frac{e^{i2k_ia_i\cos\theta_i}}{D_{ji}} \left\{\sqrt{\frac{\epsilon_i}{\mu_i}}\cos^2\theta_i\sec\theta_j\left(4\pi{\sigma}_{xx}+\sqrt{\frac{\epsilon_j}{\mu_j}}\sec\theta_j\right)-\sqrt{\frac{\epsilon_i}{\mu_i}}\left(\sqrt{\frac{\epsilon_j}{\mu_j}}+4\pi{\sigma}_{xx}\sec\theta_j\right)\right.
\nonumber \\
&&\left.+\cos\theta_i\left\{4\pi\sqrt{\frac{\epsilon_j}{\mu_j}}{\sigma}_{xx}+\sec\theta_j\left[-\frac{\epsilon_i}{\mu_i}+\frac{\epsilon_j}{\mu_j}+16\pi^2\left({\sigma}_{xx}^2+{\sigma}_{xy}^2\right)+4\pi\sqrt{\frac{\epsilon_j}{\mu_j}}{\sigma}_{xx}\sec\theta_j\right]\right\}\right\},
\nonumber \\
r_{xy} &=&
-\frac{e^{i2k_ia_i\cos\theta_i}}{D_{ji}}8\pi\sqrt{\frac{\epsilon_i}{\mu_i}}{\sigma}_{xy}\cos\theta_i\sec\theta_j, 
\label{oblique} \\
t_{xx} &=&
\frac{e^{i\left(k_i\cos\theta_i-k_j\cos\theta_j\right)a_i}}{D_{ji}}2
\sqrt{\frac{\epsilon_i}{\mu_i}}\cos\theta_i\sec\theta_j\left[\sqrt{\frac{\epsilon_i}{\mu_i}}+\cos\theta_i\left(4\pi{\sigma}_{xx}+\sqrt{\frac{\epsilon_j}{\mu_j}}\sec\theta_j\right)\right],
\nonumber \\
t_{yy} &=&
\frac{e^{i\left(k_i\cos\theta_i-k_j\cos\theta_j\right)a_i}}{D_{ji}}2
\sqrt{\frac{\epsilon_i}{\mu_i}}\cos\theta_i\sec\theta_j\left[\sqrt{\frac{\epsilon_j}{\mu_j}}+\sec\theta_j\left(4\pi{\sigma}_{xx}+\sqrt{\frac{\epsilon_i}{\mu_i}}\cos\theta_i\right)\right],
\nonumber \\
t_{xy} &=&
-\frac{e^{i\left(k_i\cos\theta_i-k_j\cos\theta_j\right)a_i}}{D_{ji}}8\pi
\sqrt{\frac{\epsilon_i}{\mu_i}}\cos\theta_i\sec\theta_j{\sigma}_{xy},
\label{obliquert}
\end{eqnarray}
\end{widetext}

\begin{widetext}
\begin{eqnarray}
r'_{xx} &=& \frac{e^{-i2k_ja_i\cos\theta_j}}{D_{ji}} \left\{-\sqrt{\frac{\epsilon_i}{\mu_i}}\cos^2\theta_i\sec\theta_j\left(4\pi{\sigma}_{xx}+\sqrt{\frac{\epsilon_j}{\mu_j}}\sec\theta_j\right)+\sqrt{\frac{\epsilon_i}{\mu_i}}\left(\sqrt{\frac{\epsilon_j}{\mu_j}}-4\pi{\sigma}_{xx}\sec\theta_j\right)\right.
\nonumber \\
&&\left.-\cos\theta_i\left\{-4\pi\sqrt{\frac{\epsilon_j}{\mu_j}}{\sigma}_{xx}+\sec\theta_j\left[\frac{\epsilon_i}{\mu_i}-\frac{\epsilon_j}{\mu_j}+16\pi^2\left({\sigma}_{xx}^2+{\sigma}_{xy}^2\right)+4\pi\sqrt{\frac{\epsilon_j}{\mu_j}}{\sigma}_{xx}\sec\theta_j\right]\right\}\right\},
\nonumber \\
r'_{yy} &=& \frac{e^{-i2k_ja_i\cos\theta_j}}{D_{ji}} \left\{-\sqrt{\frac{\epsilon_i}{\mu_i}}\cos^2\theta_i\sec\theta_j\left(4\pi{\sigma}_{xx}-\sqrt{\frac{\epsilon_j}{\mu_j}}\sec\theta_j\right)-\sqrt{\frac{\epsilon_i}{\mu_i}}\left(\sqrt{\frac{\epsilon_j}{\mu_j}}+4\pi{\sigma}_{xx}\sec\theta_j\right)\right.
\nonumber \\
&&\left.-\cos\theta_i\left\{4\pi\sqrt{\frac{\epsilon_j}{\mu_j}}{\sigma}_{xx}+\sec\theta_j\left[\frac{\epsilon_i}{\mu_i}-\frac{\epsilon_j}{\mu_j}+16\pi^2\left({\sigma}_{xx}^2+{\sigma}_{xy}^2\right)-4\pi\sqrt{\frac{\epsilon_j}{\mu_j}}{\sigma}_{xx}\sec\theta_j\right]\right\}\right\},
\nonumber \\
r'_{xy} &=&
-\frac{e^{-i2k_ja_i\cos\theta_j}}{D_{ji}}8\pi\sqrt{\frac{\epsilon_j}{\mu_j}}{\sigma}_{xy}\cos\theta_i\sec\theta_j, 
\label{oblique} \\
t'_{xx} &=&
\frac{e^{i\left(k_i\cos\theta_i-k_j\cos\theta_j\right)a_i}}{D_{ji}}2
\sqrt{\frac{\epsilon_j}{\mu_j}}\left[\sqrt{\frac{\epsilon_i}{\mu_i}}+\cos\theta_i\left(4\pi{\sigma}_{xx}+\sqrt{\frac{\epsilon_j}{\mu_j}}\sec\theta_j\right)\right],
\nonumber \\
t'_{yy} &=&
\frac{e^{i\left(k_i\cos\theta_i-k_j\cos\theta_j\right)a_i}}{D_{ji}}2
\sqrt{\frac{\epsilon_j}{\mu_j}}\left[\sqrt{\frac{\epsilon_j}{\mu_j}}+\sec\theta_j\left(4\pi{\sigma}_{xx}+\sqrt{\frac{\epsilon_i}{\mu_i}}\cos\theta_i\right)\right],
\nonumber \\
t'_{xy} &=&
-\frac{e^{i\left(k_i\cos\theta_i-k_j\cos\theta_j\right)a_i}}{D_{ji}}8\pi
\sqrt{\frac{\epsilon_j}{\mu_j}}\cos\theta_i\sec\theta_j{\sigma}_{xy},
\label{obliquertp}
\end{eqnarray}

where
\begin{eqnarray}
D_{ji} &=&
\sqrt{\frac{\epsilon_i}{\mu_i}}\cos^2\theta_i\sec\theta_j\left(4\pi{\sigma}_{xx}+\sqrt{\frac{\epsilon_j}{\mu_j}}\sec\theta_j\right)+\sqrt{\frac{\epsilon_i}{\mu_i}}\left(\sqrt{\frac{\epsilon_j}{\mu_j}}+4\pi{\sigma}_{xx}\sec\theta_j\right)
\nonumber \\
&&+\cos\theta_i\left\{4\pi\sqrt{\frac{\epsilon_j}{\mu_j}}{\sigma}_{xx}+\sec\theta_j\left[\frac{\epsilon_i}{\mu_i}+\frac{\epsilon_j}{\mu_j}+16\pi^2\left({\sigma}_{xx}^2+{\sigma}_{xy}^2\right)+4\pi\sqrt{\frac{\epsilon_j}{\mu_j}}{\sigma}_{xx}\sec\theta_j\right]\right\}.
\label{Den}
\end{eqnarray}
\end{widetext}
Note that $r_{yy}, t_{yy}$ are no longer equal to  $r_{xx},
t_{xx}$ at oblique light incidence. Eqs.~(\ref{obliquert})-(\ref{Den}) recover the normal
  incidence results Eqs.~(\ref{normalr})-(\ref{normalt}) when
  $\theta_i = \theta_j = 0$.

For completeness, we also include the expressions of the total reflection and
transmission tensors in the presence of a layer of dielectric
substrate. These can be composed from the expressions Eqs.~(\ref{r20})-(\ref{t20}):
\begin{eqnarray}
\bar{r} &=& 
\bar{r_{\mathrm{T}}}+\bar{t_{\mathrm{T}}'}\bar{r_{\mathrm{S,B}}}\left(\bm{1}-\bar{r_{\mathrm{T}}'}\bar{r_{\mathrm{S,B}}}\right)^{-1}\bar{t_{\mathrm{T}}}, \label{r30} \\
\bar{t} &=&
\bar{t_{\mathrm{S,B}}}\left(\bm{1}-\bar{r_{\mathrm{T}}'}\bar{r_{\mathrm{S,B}}}\right)^{-1}\bar{t_{\mathrm{T}}}. \label{t30}
\end{eqnarray}
where $\bar{r_{\mathrm{S,B}}^{(')}}$ and $\bar{t_{\mathrm{S,B}}^{(')}}$ (the
subscript 'S' denotes substrate) are the 
reflection and transmission tensors for light propagation from the 
bottom surface to the substrate-vacuum interface 
\begin{eqnarray}
\bar{r_{\mathrm{S,B}}} &=& 
\bar{r_{\mathrm{B}}}+\bar{t_{\mathrm{B}}'}\bar{r_{\mathrm{S}}}\left(\bm{1}-\bar{r_{\mathrm{B}}'}\bar{r_{\mathrm{S}}}\right)^{-1}\bar{t_{\mathrm{B}}}, \label{r32} \\
\bar{t_{\mathrm{S,B}}} &=&
\bar{t_{\mathrm{S}}}\left(\bm{1}-\bar{r_{\mathrm{B}}'}\bar{r_{\mathrm{S}}}\right)^{-1}\bar{t_{\mathrm{B}}}, \label{t32}
\end{eqnarray}
where $\bar{r_{\mathrm{S}}^{(')}}$ and $\bar{t_{\mathrm{S}}^{(')}}$
are the reflection and transmission tensors for light scattering at the substrate-vacuum
interface.

\end{document}